\documentclass[aps,preprintnumbers,superscriptaddress,longbibliography,amsmath,amssymb,twocolumn,10pt,floatfix]{revtex4-2}

\usepackage{graphicx}
\usepackage{dcolumn}
\usepackage{bm}
\usepackage{xspace}
\usepackage{multirow}
\usepackage{booktabs}
\usepackage{xcolor}
\usepackage{xfrac}
\usepackage[caption=false]{subfig}
\graphicspath{{./plots/}}

\usepackage{hyperref}
\hypersetup{
    colorlinks=true,
    linkcolor=blue,
    filecolor=blue,      
    urlcolor=blue,
    pdftitle={Overleaf Example},
    pdfpagemode=FullScreen,
    }

\usepackage{units}
\usepackage{comment}

\newcommand{\x}[1]{%
  {}$
  \kern-2\mathsurround 
  $
  \binoppenalty10000 \relpenalty10000 #1
  {}$
  \kern-2\mathsurround 
  $
}

\begin{document}


\title{Discriminating sub-TeV gamma and hadron-induced showers through their footprints}
\date{\today}

\author{R. Concei\c{c}\~{a}o}
\affiliation{Laborat\'orio de Instrumenta\c{c}\~ao e F\'\i{}sica Experimental de Part\'\i{}culas -- LIP and Instituto Superior T\'ecnico -- IST, Universidade de Lisboa -- UL, Lisboa, Portugal}

\author{B.~S.~Gonz\'alez}
\email{borjasg@lip.pt}
\affiliation{Laborat\'orio de Instrumenta\c{c}\~ao e F\'\i{}sica Experimental de Part\'\i{}culas -- LIP and Instituto Superior T\'ecnico -- IST, Universidade de Lisboa -- UL, Lisboa, Portugal}

\author{A. Guill\'en}
\affiliation{Computer Engineering, Automatics and Robotics Department, University of Granada, Granada, Spain}

\author{M.~Pimenta}
\affiliation{Laborat\'orio de Instrumenta\c{c}\~ao e F\'\i{}sica Experimental de Part\'\i{}culas -- LIP and Instituto Superior T\'ecnico -- IST, Universidade de Lisboa -- UL, Lisboa, Portugal}

\author{B.~Tom\'e}
\affiliation{Laborat\'orio de Instrumenta\c{c}\~ao e F\'\i{}sica Experimental de Part\'\i{}culas -- LIP and Instituto Superior T\'ecnico -- IST, Universidade de Lisboa -- UL, Lisboa, Portugal}

\begin{abstract}
    Gamma/hadron discrimination in ground-based gamma-ray observatories at the sub-TeV energy range is challenging as traditional muon-based methods become less effective at lower energies. This work explores a novel gamma/hadron discrimination method for Extensive Air Shower arrays that analyzes the shower signal footprint patterns using a state-of-the-art pre-trained Vision Transformer (ViT). The resilience of the method to background noise, such as atmospheric muons and low-energy proton showers, along with its adaptability to different zenith angles and array configurations, demonstrates its potential for application in current and future ground-based gamma-ray observatories.
\end{abstract}

\pacs{Valid PACS appear here}
\maketitle


\section{Introduction}
\label{sec:intro}
The study of gamma rays is crucial for understanding extreme astrophysical events and probing fundamental physics. Gamma-rays in the sub-TeV to TeV range can provide insights into Active Galactic Nuclei (AGNs), Gamma Ray Bursts (GRBs), and potential new physics beyond the standard model, including dark matter research \cite{Pimenta2018Astroparticle,viana2019searching}.

Their detection above a few hundred of GeV is possible using ground-based detectors as satellite detection is limited by low fluxes \cite{atwood2009large}. These experiments indirectly detect gamma rays at very high altitudes by using Extensive Air Showers (EAS), which are a cascade of secondary shower particles produced in the atmosphere due to the interaction of the gamma-ray with it. Then, the characteristics of the primary particle such as energy and direction are inferred from the secondary particles detected at the ground. This method enables surveying large portions of the sky and is sensitive to transient phenomena. However, gamma rays must be distinguished from the dominant hadronic background \cite{degrange2015introduction, albert2019science,torres2024enhanced}. 

Muon identification remains one of the main methods for gamma/hadron separation at TeV energies \cite{wcd2022mercedes,WCD4PMTs,4PMTs_NCA,WCD_9iPM_2020,kunwar2023double}. However, the muon content of hadronic showers is scarce at sub-TeV energies. Alternatively, one may analyze the shower footprint patterns on the ground to infer features of the shower development \cite{ARGO_2013,LCm2022gamma,Ptail2024,HAWC_Crab}. Hadron-induced showers, unlike pure electromagnetic showers, generate particles with high transverse momentum, causing the shower to spread more laterally and form clusters. Initial studies have shown great potential in using these ground patterns for gamma/hadron discrimination, but further work is needed to fully understand and optimize their effectiveness against noise \cite{assunccao2019automatic,gnn2024jonas}.

Machine learning (ML) has recently become an innovative instrument in the field of physics, especially within astroparticle physics, enabling advancements in numerous domains that require extensive data analysis. In modern cosmic and gamma-ray observatories, ML techniques have shown significant potential to enhance gamma/hadron separation \cite{Shilon2019, jonas_gnn_IACT_2023, ghsep_2024_tibet,lhaaso_ghsep_2020, LHAASO2021performance,LHAASO_2020_gnn,HAWC2021ML}, event reconstruction \cite{carrillo2019improving,GUILLEN201912,nieto2019ctlearn,erdmann2018deep}, and even allow neutrino identification with water Cherenkov detectors (WCDs) \cite{neutrinos_mercedes_2024}. These breakthroughs are analogous to significant advances in related domains, such as the IceCube detection of neutrinos from the galactic plane \cite{icecube2023galacticplane}, investigations on the mass composition and $X_{\rm max}$ estimation of ultra-high-energy cosmic rays (UHECRs) by the Pierre Auger Observatory \cite{auger2021xmax,Mass2023Auger,auger2024mass}, and the search of new physics with the Large Hadron Collider (LHC) \cite{LHC_2018, LHC2020deep, LHC2022deep, apolinario2021deep, deeplearning2015nature}.

Among ML-based techniques, transformers, which use attention mechanisms for data analysis \cite{attention_2017}, have emerged as particularly effective. Their initial applications in event reconstruction for astroparticle experiments have demonstrated promising results \cite{Watson2023HAWC}. In this work, we demonstrate that state-of-the-art pre-trained Vision Transformers (ViTs) \cite{ViT_2020} have significant potential to accurately discriminate between gamma and hadron-induced air showers. The footprint image, created from signals detected by the individual WCDs in the detector array, is used as input for the model. The method's robustness against noise, including atmospheric muons and low-energy proton showers, is also assessed.

The paper is structured as follows: Section \ref{sec:simulations} describes the set of simulations used and the detector configuration. Section \ref{sec:analysis} details the analysis of shower footprint patterns. Section \ref{sec:results} evaluates the method's performance and resilience to various noise sources. Finally, the conclusions are presented in Section \ref{sec:conclusions}.

\section{Set of simulations}\label{sec:simulations}
Following the same simulation strategy as in \cite{WCD4PMTs,wcd2022mercedes}, the Extensive Air Showers were simulated with CORSIKA (version 7.5600) \cite{CORSIKA}, taking FLUKA \cite{fluka_2005,fluka_2014} and QGSJet-II.04 \cite{QGSJET_2011} as the low- and high-energy hadronic interaction model, respectively. The detector response was simulated with the Geant4 toolkit (version 4.10.05.p01) \cite{agostinelli2003geant4,Geant4_2006,Geant4_2016}.

The experimental observation level was established at $5\,200$ m above sea level \footnote{This altitude corresponds to the altitude of the ALMA site in Chile.}. The detector array comprises $5\,720$ Mercedes WCDs \cite{wcd2022mercedes}, covering an area of $80\,000 \, \rm m^2$ with a fill factor of approximately $85\%$. Two additional arrays, with fill factors of $65\%$ and $40\%$, were generated from the initial simulation through the uniform removal of stations, as explained in Subsection \ref{sec:fill_factor} (see Figure \ref{fig:fill_factor}).

The CORSIKA simulations were generated using a \( E^{-1} \) spectrum to balance computational time with statistical precision. During the analysis, event weights were applied to correct the spectra of gamma rays and protons to spectral indices of \( E^{-2} \) and \( E^{-3} \), respectively, to ensure a realistic power-law energy spectrum of cosmic rays. 

The showers are nearly vertical, with simulated zenith angles of $\theta_0 = 10^{\circ}$ and $\theta_0 \in \left[5^{\circ}; 15^{\circ}\right]$ for gammas and protons, respectively. To evaluate the effectiveness of the method for showers with greater inclination, simulations were also conducted for gamma and proton events at zenith angles of $\theta_0 = 30^{\circ}$ and within the range $\theta_0 \in \left[25^{\circ}; 35^{\circ}\right]$, respectively. The choice of these wider zenith angle ranges for protons reflects the diffuse nature of the background, as a realistic scenario would involve direction reconstruction and accept events with slightly different zenith angles. Nonetheless, as discussed in Subsection \ref{sec:zenith_angle}, the performance of the ViT exhibits minimal dependence on the zenith angle.

Each CORSIKA shower was simulated five times using a Geant4 dedicated simulation with different core positions to enhance statistical reliability. The simulated cores were uniformly distributed within a radius of $165 \, \rm m$, slightly exceeding the radius of the array of $160 \, \rm m$, and their impact on the gamma/hadron discrimination power is assessed in Subsection~\ref{sec:shower_core}.
The event was simulated under four conditions: without noise, with noise from atmospheric muons, noise from low-energy proton showers, and noise from both sources. The data set is split such that a given shower appears in only one dataset, regardless of its core position or contamination. This approach ensures realistic results, preventing data leakage, where the model could learn a specific shower in training and test it correctly in other sets. It also helps the network learn that the shower corresponds to a given primary particle, regardless of contamination or core position.

\begin{figure}[htb]
 \centering
\includegraphics[width=0.99\linewidth]{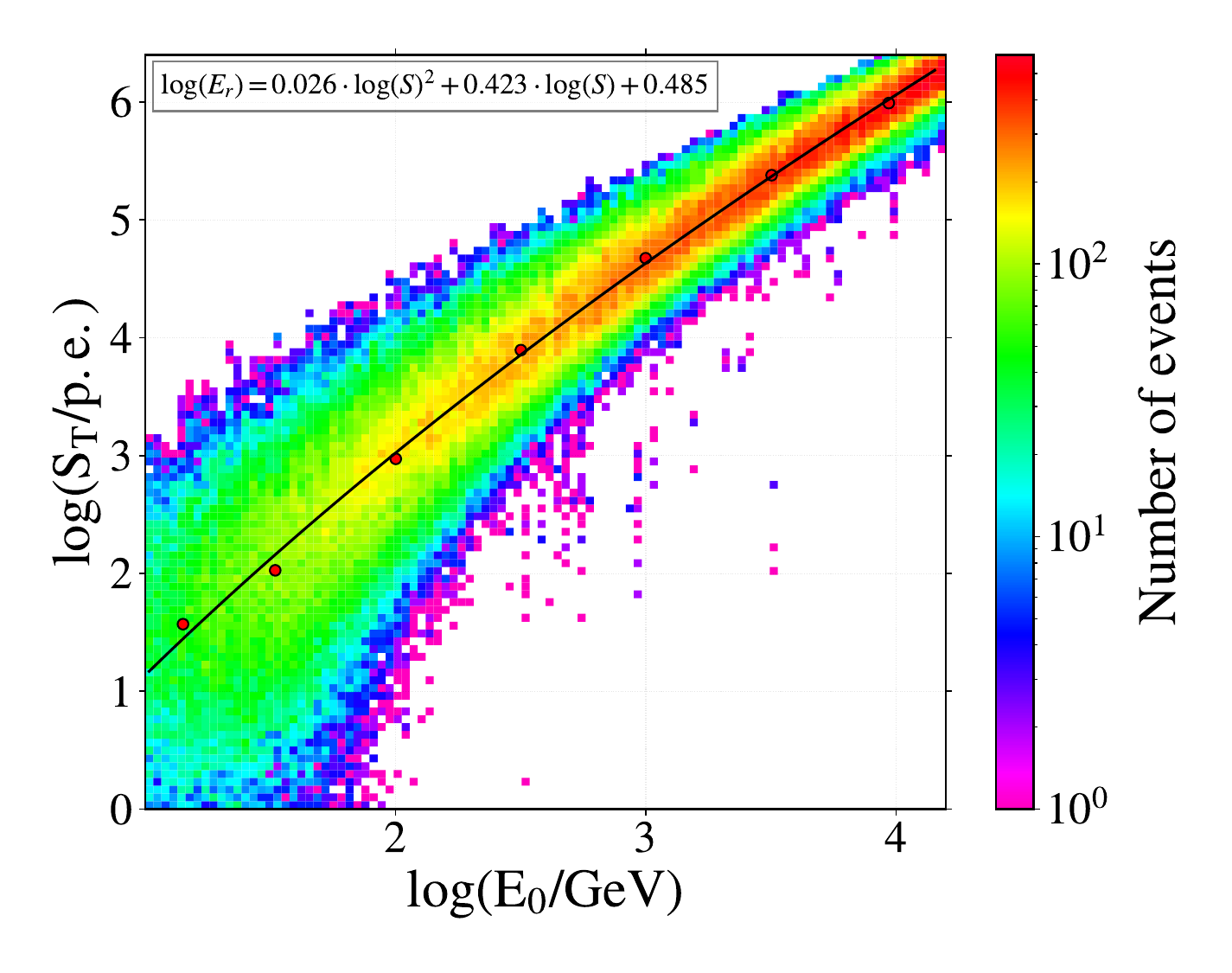}
\caption{Energy calibration using gamma-induced showers with $\theta_0=10^\circ$ and $r_{core} \in \left[0;165\right] \, \rm m$. The black line represents the calibration and the red points the mean signal for each energy bin.}
\label{fig:energy_calibration} 
\end{figure}

\begin{figure}[htb]
 \centering
\includegraphics[width=0.99\linewidth]{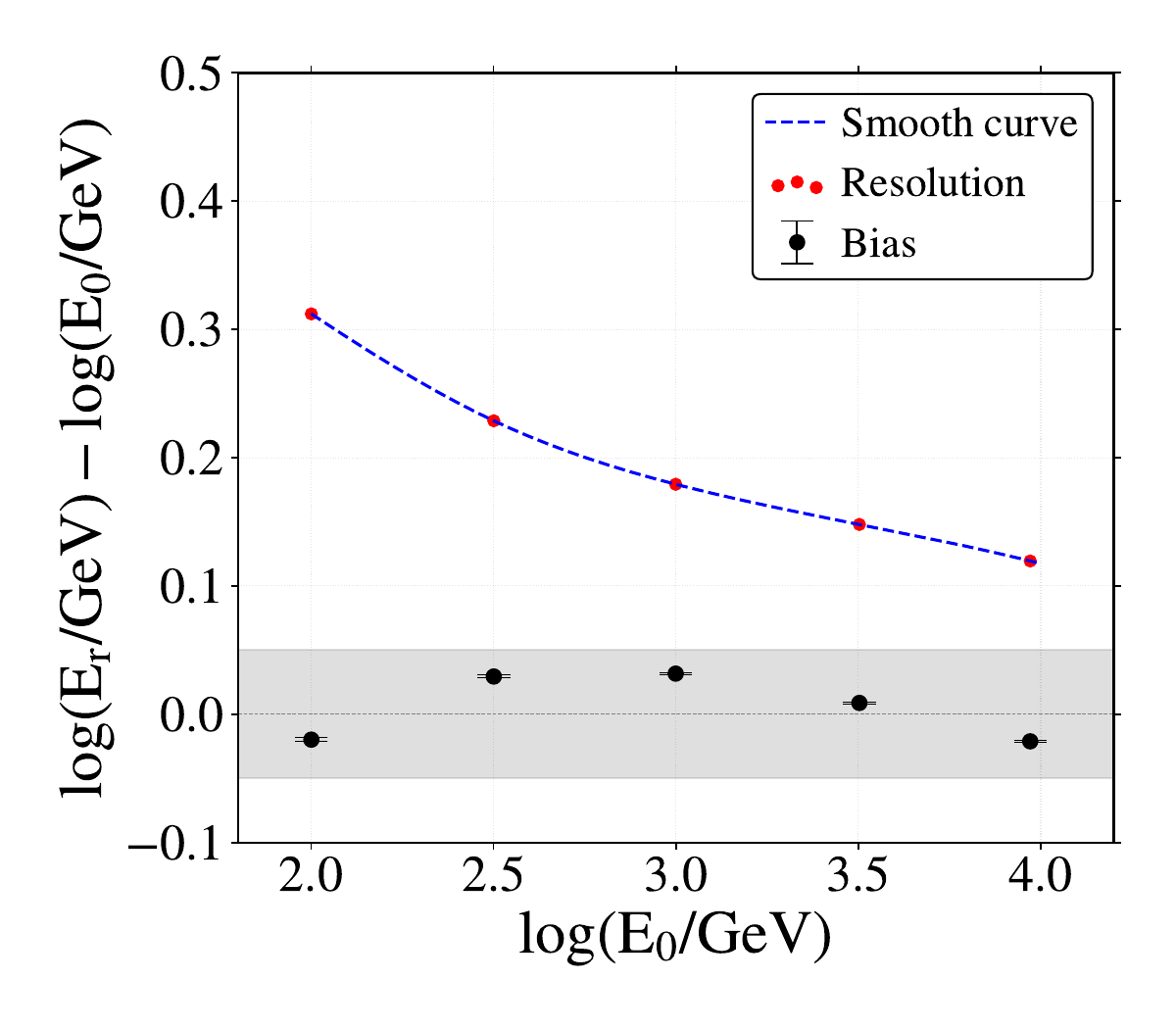}
\caption{Bias and resolution of the energy calibration. The grey area highlights the region with a bias of $\leq 5\%$. A smooth curve was incorporated to guide the eye.
}
\label{fig:energy_calibration_bias} 
\end{figure}

Three energy bins were selected to evaluate this method, corresponding to reconstructed energies of approximately $200 \, \rm{GeV}$, $500 \, \rm{GeV}$, and $1 \, \rm{TeV}$. A calibration was performed using the total signal at the ground to emulate the typical energy reconstruction in ground-based observatories, as shown in Figure \ref{fig:energy_calibration}. The energy bins were chosen according to the energy resolution of the selected bin (see Figure \ref{fig:energy_calibration_bias}), which is about $50 \, \%$ at 500 GeV. Thus, the analysis bins are defined as $E_r \in \left[200, 500\right]\, \rm{GeV}$, $E_r \in \left[500, 900\right]\, \rm{GeV}$, and $E_r \in \left[1\,000, 1\,600\right]\, \rm{GeV}$. To guarantee realistic results, simulations were performed for primaries with energies ranging from $10 \, \rm GeV$ to $10 \, \rm TeV$ (see Figure \ref{fig:energy_calibration}).

The number of events generated after having simulated five core positions for each shower and adding the above-mentioned contamination sources are detailed for each energy bin in Table \ref{tab:datasets}. The data set was divided into training, validation, and test sets, where the training set is used to train the model, the validation set is used to tune hyperparameters and assess performance during development, and the test set is reserved for evaluating the final model performance.

\begin{table}[htb]
    \centering
    \begin{tabular}{|c|c|c|c|c|c|}
        \hline
        \textbf{Energy} & \(\boldsymbol{\theta}\) & \textbf{Split} & \textbf{Gammas} & \textbf{Protons} & \textbf{Total} \\
        \hline
        \multirow{3}{*}{200 GeV} 
        & $10^\circ$ & Train & $79\,747$ & $77\,357$ & $157\,104$ \\
        & $10^\circ$ & Test & $22\,582$ & $22\,791$ & $45\,373$ \\
        & $10^\circ$ & Validation & $13\,787$ & $12\,703$ & $26\,490$ \\
        & $30^\circ$ & Test & $1\,563$ & $1\,286$ & $2\,849$ \\
        \hline
        \multirow{3}{*}{500 GeV} 
        & $10^\circ$ & Train & $54\,947$ & $57\,022$ & $111\,969$ \\
        & $10^\circ$ & Test & $15\,790$ & $16\,525$ & $32\,315$ \\
        & $10^\circ$ & Validation & $8\,883$ & $9\,502$ & $18\,385$ \\
        & $30^\circ$ & Test & $1\,208$ & $968$ & $2\,176$ \\
        \hline
        \multirow{3}{*}{1 TeV} 
        & $10^\circ$ & Train & $43\,669$ & $44\,532$ & $88\,201$ \\
        & $10^\circ$ & Test & $12\,447$ & $12\,612$ & $25\,059$ \\
        & $10^\circ$ & Validation & $7\,030$ & $7\,394$ & $14\,424$ \\
        & $30^\circ$ & Test & $899$ & $584$ & $1\,483$ \\
        \hline
    \end{tabular}
    \caption{Number of EAS events in each dataset.}
    \label{tab:datasets}
\end{table}

\begin{figure*}[t!]
 \centering
  \subfloat[Gamma-induced shower event.]{
   \label{fig:photon_example}
    \frame{\includegraphics[width=0.33\textwidth]{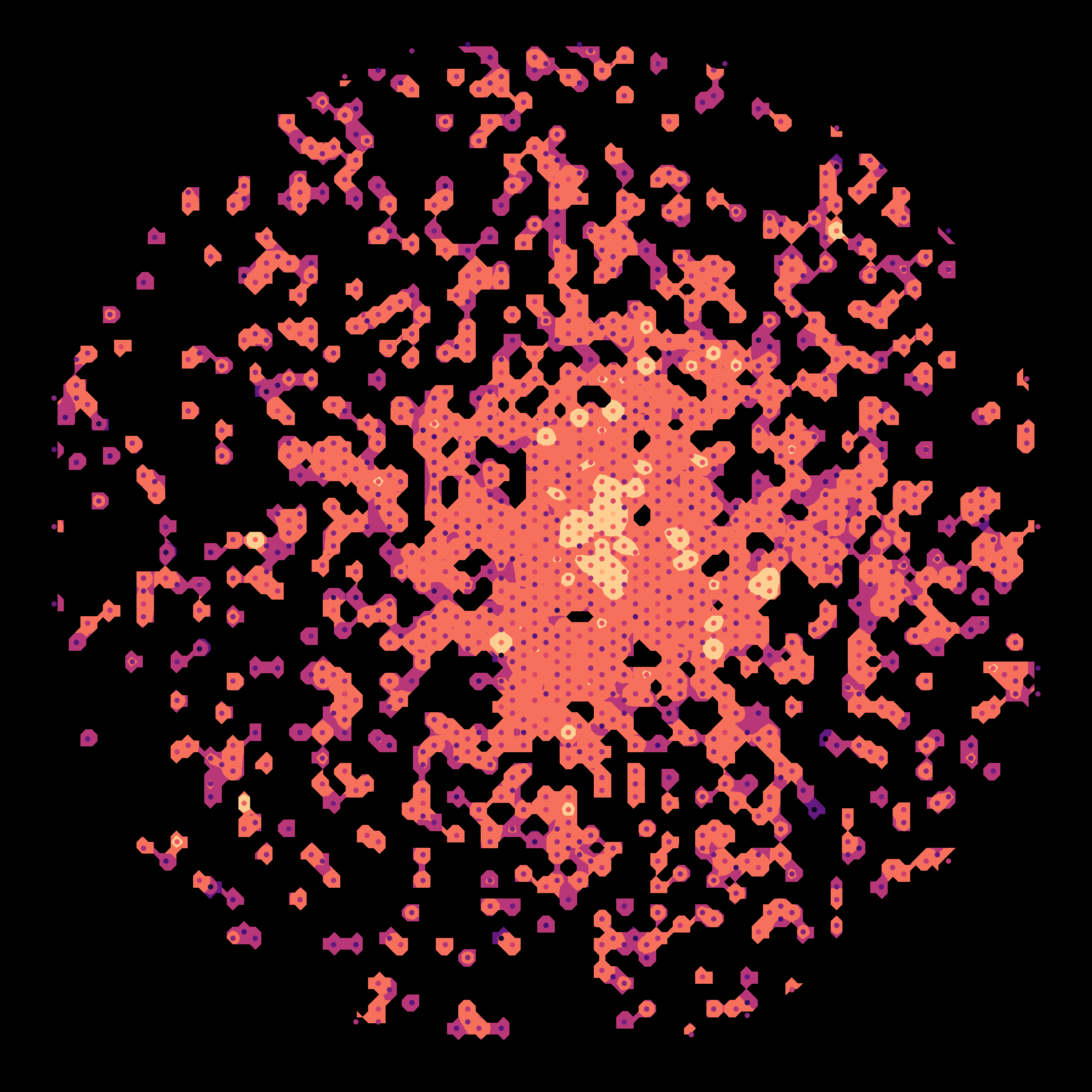}}}
    \hspace{0.1in}
    \subfloat[Proton-induced shower event.]{
 \label{fig:proton_example}
    \frame{\includegraphics[width=0.33\textwidth]{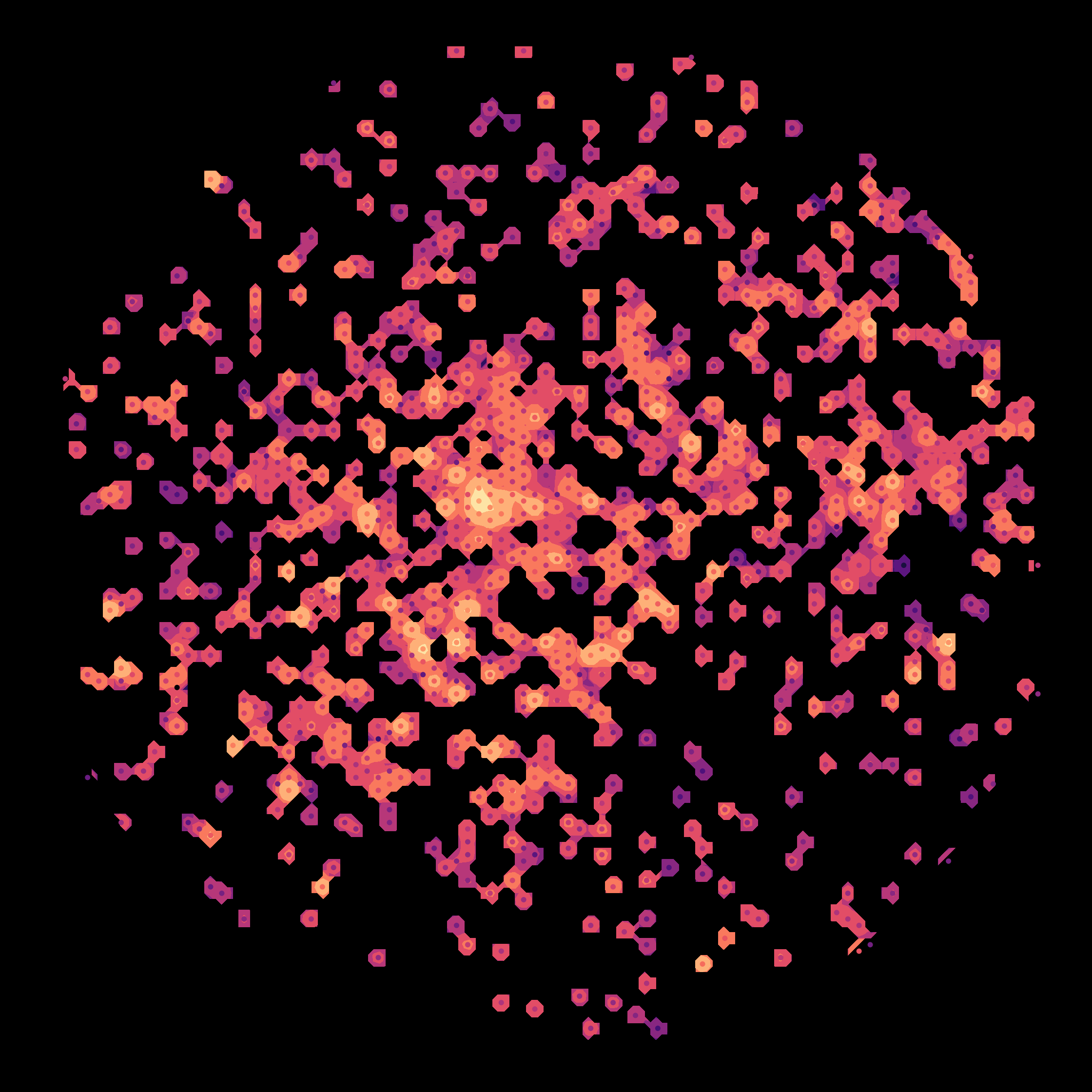}}}
 \caption{Shower events with similar reconstructed energy ($E_r \sim 1 \, \rm{TeV}$) and shower core position. The station color scale ranges from $0 \, \rm p.e.$ (black) to $10^5 \, \rm p.e.$ (brightest color). The contour groups the stations into five color levels, from the lowest (darkest) to the highest (brightest) signals.
}
 \label{fig:example_input}
\end{figure*}

\section{Analysis strategy: Vision Transformers}\label{sec:analysis}
ViTs are emerging in the field of computer vision. Unlike traditional Convolutional Neural Networks (CNNs) \cite{CNNs_2015_introduction}, which rely on convolutional and fully connected layers, ViTs depend on self-attention mechanisms, enabling them to capture long-range dependencies and context within images \cite{attention_2017}. To do this, the image is divided into fixed-size non-overlapping patches, each of which is linearly embedded, followed by the addition of position embeddings. Subsequently, the resulting sequence of vectors is fed into a standard Transformer encoder, where extra learnable classification tokens are added for classification. This innovation enables these models to capture intricate long-range dependencies and contextual information within images, leading to enhanced capabilities in various applications. 

In this study, the state-of-the-art pre-trained ViT model for classification ``\textit{google/vit-base-patch16-224-in21k}'' \cite{ViT_2020,vit-base-patch16-224-in21k,wu2020visual,deng2009imagenet} has been used. This model is a BERT-like transformer encoder that includes an extra linear layer applied to the classification token ([CLS]) for executing classification tasks. It was pre-trained on a large data set called \textit{ImageNet-21k} \cite{Imagenet-21k_2021}, which is composed of 14 million images at a resolution of $224\times224$ pixels and has $21\,843$ classes. It is crucial to note that using a pre-trained ViT considerably lowers the computational expense involved in training from scratch, while taking advantage of the rich feature representations obtained from large data sets. For the sake of reproducibility, we employed the ViT model available at the following URL: \url{https://huggingface.co/docs/transformers/model_doc/vit#transformers.ViTForImageClassification}.

Before feeding the ViT model with images, they are divided into a sequence of fixed-size non-overlapping patches. It is important to note that the transformer sequence length is inversely proportional to the square of the patch size. Consequently, while smaller patches allow for more detailed image analysis, they also increase computational complexity \cite{ViT_2020}. As a compromise between computational efficiency and array area covered, a patch size of $16\times16$ pixels was selected, corresponding to an actual area of $25\, \rm m \times 25\, \rm m$.

The model was trained on our problem data for 30 additional epochs using a batch size of 32, the default learning rate of $5 \times 10^{-5}$, a warm-up ratio of $10\%$, and the mean square loss function. This pipeline was done with the \textit{Hugging Face Transformers} open-source library, which provides APIs and tools to download and train state-of-the-art pre-trained models \cite{huggingface_2019}. 

The input of the model is an image created by mapping the positions of the WCDs and assigning colors based on the total signal in each tank. Based on the optimal efficiency attained during the experimental stage, the Matplotlib \cite{matplotlib2007} \textit{Magma} colormap was utilised, with a dark background selected to depict the absence of energy in those areas. To highlight signal patterns within the footprint and cluster stations with comparable signals, a contour plot fills the space between the WCDs, using a grid with cells similar in size to the stations. This contour plot represents the logarithmic signal intensity levels, spaced over the range of observed signals, employing five equally spaced levels to illustrate gradients in signal distribution. The chosen number of levels balances the need to group stations effectively without over-segmenting the signals. Although fewer levels fail to capture meaningful groupings, higher numbers introduce unnecessary segmentation. The test with twenty levels yielded comparable results, supporting the suitability of five levels for this analysis. The continuous signal range of the WCDs and the discrete contour use an identical colormap to highlight stations with marginally higher or lower signals within a group. A logarithmic scale is applied to prevent saturation of the color scale due to the high signals in the stations near the shower core. 

To reduce the noise of background or late-arriving particles, the shower plane is fitted using the positions of all active stations, with the $T_0$ of each station as the Z coordinate~\cite{trigger_paper}. The $T_0$ for each station is defined as the timing bin within the signal trace where it records a minimum of 3 photoelectrons across all PMTs, with time bins of $1 \, \rm ns$ \cite{wcd2022mercedes}. It was found that with an average background of 23 stations, the plane fit successfully converges when at least 30 stations are present in the event. Therefore, only events with more than 30 active stations (signal + background) were considered for analysis. Following this procedure, only stations with a temporal offset of less than $50\, \rm ns$ from the fitted shower plane were selected. By applying the shower plane fit, approximately $40\%$ of the noise stations are excluded, while only about $20\%$ of the signal stations are eliminated. It should be noted that most of the excluded signal stations are those arriving late and not correlated with the shower plane, potentially lacking additional contribution to the footprint information. Thus, while identical results are observed for noise-free events, the approach effectively reduces the degradation caused by noise.

In addition, as part of the preprocessing steps, the images are rescaled to the same resolution ($224\times224$) and normalised across the RGB channels with mean (0.5, 0.5, 0.5) and standard deviation (0.5, 0.5, 0.5) \cite{vit-base-patch16-224-in21k}. 

The outcome of this preprocessing pipeline is illustrated in Figure \ref{fig:example_input} for proton and gamma-ray shower events, each with a reconstructed energy of 1 TeV. Certain distinguishing features of each class can be discerned from these examples. The gamma-ray event displays a well-defined shower core with a uniform signal, characteristic of a purely electromagnetic shower. In contrast, the proton event shows a more irregular and complex footprint, with clusters of signal appearing far from the shower core, resulting from particles with high transverse momentum generated in hadronic interactions.

Finally, the output of the ViT is defined as a classification probability, denoted as $P \in \left[0;1\right]$, that a proton has induced the analyzed shower.

\begin{figure*}[htb]
 \centering
  \subfloat[Noise free.]{
   \label{fig:photon_example_noise_free}
    \frame{\includegraphics[width=0.3\textwidth]{Fig4a.pdf}}}
    \hspace{0.05in}
    \subfloat[Contamination by atmospheric muons with 29 triggered stations and a total signal of $6\,600$ p.e.]{
 \label{fig:photon_example_muon_noise}
    \frame{\includegraphics[width=0.3\textwidth]{Fig4b.pdf}}}\hspace{0.05in}
    \subfloat[Contamination with a 126 GeV proton-induced shower with 68 triggered stations and a total signal of $2\,300$ p.e.]{
 \label{fig:photon_example_shower_noise}
    \frame{\includegraphics[width=0.3\textwidth]{Fig4c.pdf}}}
 \caption{Example of contamination for a photon-induced shower event with a reconstructed energy of $550$ GeV.
}
 \label{fig:example_shower_noise}
\end{figure*}

\section{Results}
\label{sec:results}
This section assesses the gamma/hadron separation capability of the method by calculating the background rejection factor, $1-\varepsilon_p$, at a signal selection efficiency of $\varepsilon_{\gamma} \sim80\%$. This value was chosen to ensure the identification of the majority of gamma-induced events. Moreover, the Appendix \ref{sec:optimal_q} provides a description of the background rejection factor corresponding to the optimised gamma efficiency. This optimisation aims to maximise the Q-factor, defined as the ratio of signal selection to the square root of background rejection. A ViT model was trained for each energy bin and tested under various noise conditions, as detailed later. 

Subsection E presents a realistic comparison of various gamma/hadron separators developed by the HAWC experiment, evaluated under the same experimental conditions.

\subsection{Resilience to Noise from Atmospheric Muons and Low-Energy Proton Showers} \label{subsec:resilience}

For an EAS array with the given characteristics, the number of stations triggered by atmospheric muons within a $200\,$ns time window is expected to follow a Poisson distribution with a mean of 23 stations~\cite{trigger_paper}. To assess the resilience of the method against noise from atmospheric muons, each shower event is contaminated with a random number of stations based on this distribution. The muonic signal added to the contaminated tanks is randomly selected using a probability function derived from the signal distribution of approximately $150\,000$ single muon events from proton showers in the $\sim 1 \, \rm TeV$ energy bin. 

Additionally, to evaluate the resilience against background showers, the events were contaminated with a single low-energy proton shower with $E_0 \in \left[10,160\right] \, \rm GeV$. The background event is selected based on a probability function that follows the proton energy spectrum. The shower is then injected at a random position within the array area, and the signals from both showers are combined.

An example of these two sources of background is depicted in Figure \ref{fig:example_shower_noise}. It can be seen that atmospheric muons generate a uniform noise with a characteristic signal across the entire footprint, while the low-energy background shower triggers stations within a small and concentrated portion of the array. Even though these background footprints may seem small, they could be potentially problematic since they can mimic the clusters and muons found in hadronic showers. 

For each chosen reconstructed energy bin, we train and test a ViT model with both the original shower-derived images and additional images featuring contamination. This method considers the substantial variation in the shower footprint across different energy bins (see Figures \ref{fig:photon_example} and \ref{fig:photon_example_noise_free}), which could otherwise lead to ambiguity without providing explicit energy information to the transformer. For instance, the footprint of low-energy gamma rays, often exhibiting dispersed signals and unclear cores, could be misclassified as hadronic showers. A unified model covering all energy bins would require integrating reconstructed energy into both training and inference, necessitating architectural modifications beyond the scope of this work. Instead, our focus remains on demonstrating the viability of the footprint-based technique across multiple energy bins, with an emphasis on the low-energy regime relevant to gamma-ray observatories.

\begin{figure}[htb]
 \centering
\includegraphics[width=0.95\linewidth]{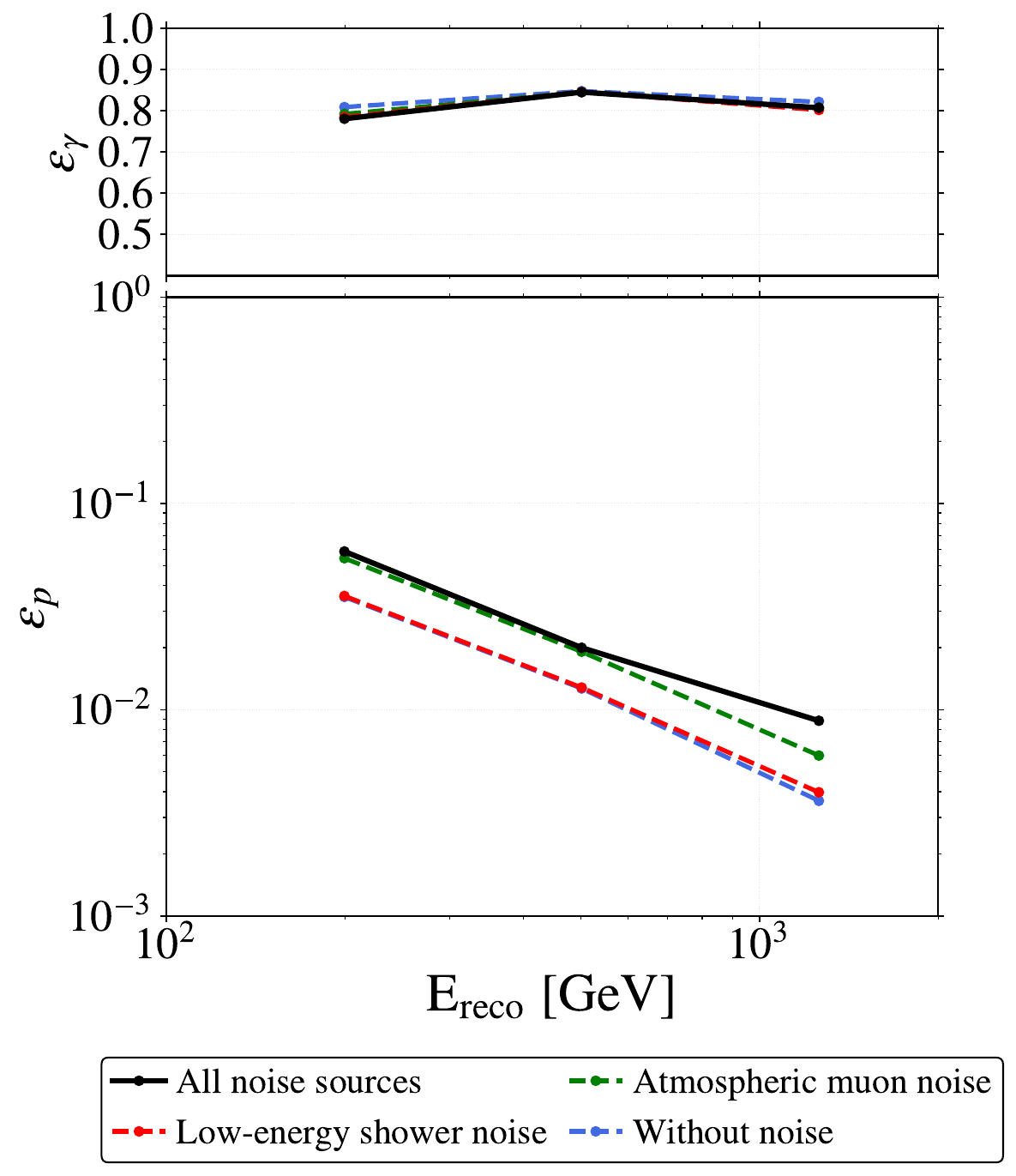}
\caption{The gamma (top) and proton (bottom) selection efficiency at different energies and noise sources.}
\label{fig:noise_results} 
\end{figure}

Figure \ref{fig:noise_results} presents the background rejection efficiency ($1-\varepsilon_p$) for the test data set, indicating that the addition of noise from low-energy proton showers has a minimal effect on the gamma/hadron separation technique, with only slight performance variations across the energy bins. However, as the energy increases, the degradation resulting from the overlap of both atmospheric muons and low-energy proton showers becomes more pronounced. This trend is expected to be related to the fact that low-energy showers rarely contain muons or signal clusters, which are similar to the noise, causing the model to rely less on those features for classification. 

Overall, the method shows exceptional background rejection across all tested scenarios.
This result demonstrates that there is enough information in the shower footprint to perform an outstanding gamma/hadron separation. 

\subsection{Performance for different zenith angles} \label{sec:zenith_angle}

A dataset consisting of gamma-ray showers at a zenith angle of $\theta_0 = 30^{\circ}$ and protons within the range $\theta_0 \in [25^{\circ};35^{\circ}]$ was simulated to assess the method's effectiveness for inclined events. For a fair comparison, the simulations were conducted using the same methodology as with vertical events: the distribution of the shower core was homogeneous in the detector area and the energy reconstruction strategy used in Figures \ref{fig:energy_calibration} and \ref{fig:energy_calibration_bias} was performed for this particular zenith angle bin. 

The background noise is independent of the primary particle; thus, the same background noise and degradation are expected for other zenith angles and only events without contamination have been used for comparison. Since a large number of events are required to train these algorithms, the model trained with vertical events in Subsection \ref{subsec:resilience} was employed for the comparison. 

\begin{figure}[htb]
 \centering
\includegraphics[width=0.95\linewidth]{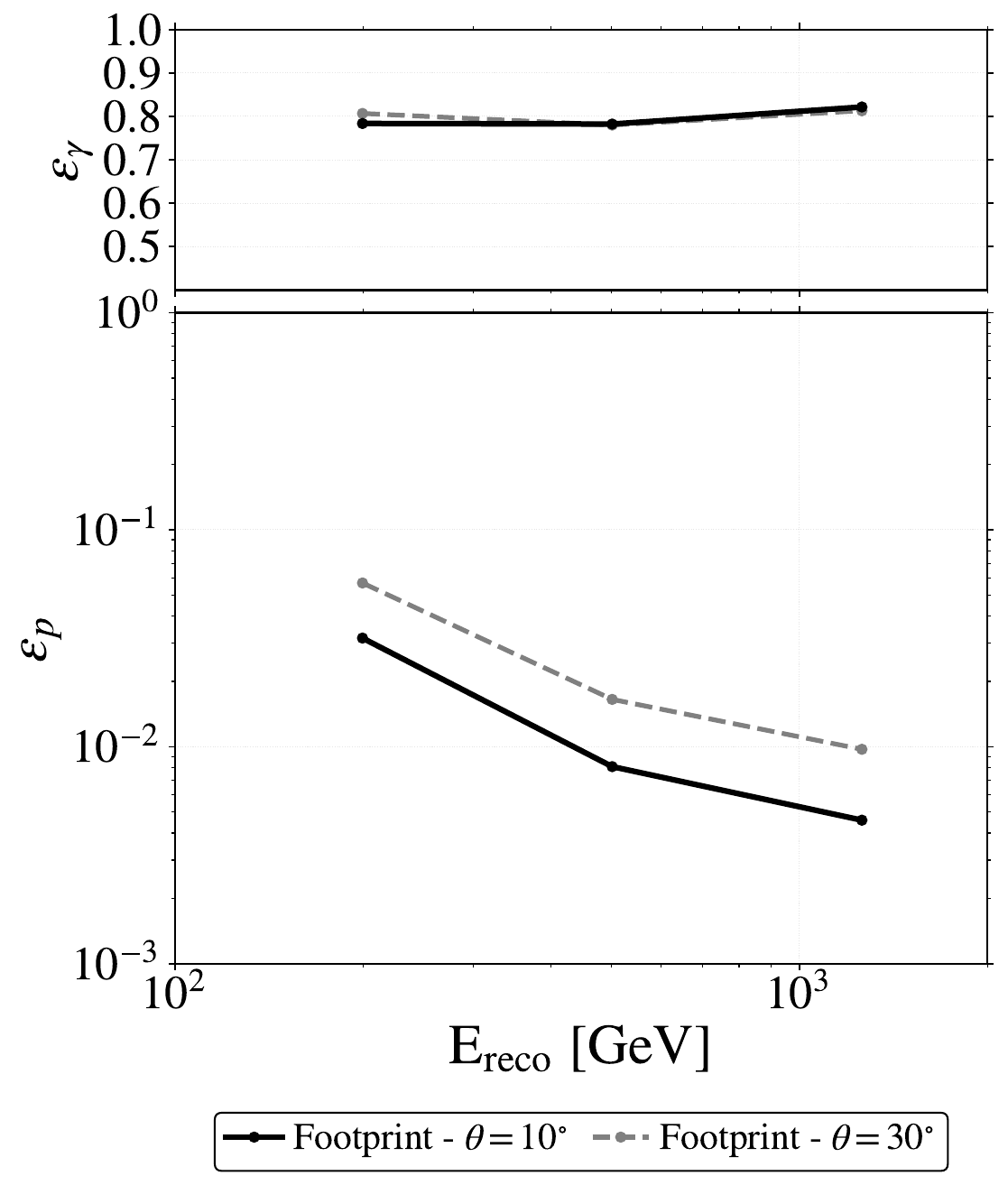}
\caption{Dependence of the gamma (top) and proton (bottom) selection efficiency on the primary zenith angle at different energy bins for shower events without noise.}
\label{fig:eff_zenith_angle} 
\end{figure}

Figure \ref{fig:eff_zenith_angle} presents the proton rejection efficiency ($1-\varepsilon_p$) attained by ViT models for shower events with zenith angles of approximately $\theta_0 \sim 10^{\circ}$ and $\theta_0 \sim 30^{\circ}$. A similar level of degradation was observed across all energy ranges, indicating overall robust performance for inclined events. It should be mentioned that further improvements might be achievable by training the network with specific angle bins, analogous to the approach taken for energy. However, implementing an angular reconstruction method is beyond the scope of this study.

\subsection{Dependence on the position of the shower core} \label{sec:shower_core}

To evaluate how the performance of the gamma/hadron discrimination model is influenced by the position of the shower core, we examined the model's output probabilities across different core positions, including those from off-array events.

To achieve this, the test shower events were iteratively simulated with their cores positioned at specific distances from the array center: $r_{\rm core} = \left\{0;50;100;150;170;180;190;200\right\} \, \rm m$. The models detailed in Subsection \ref{subsec:resilience}, trained exclusively on on-array events, were used for the evaluation.

\begin{figure}[htb]
  \centering
 \includegraphics[width=0.95\linewidth]{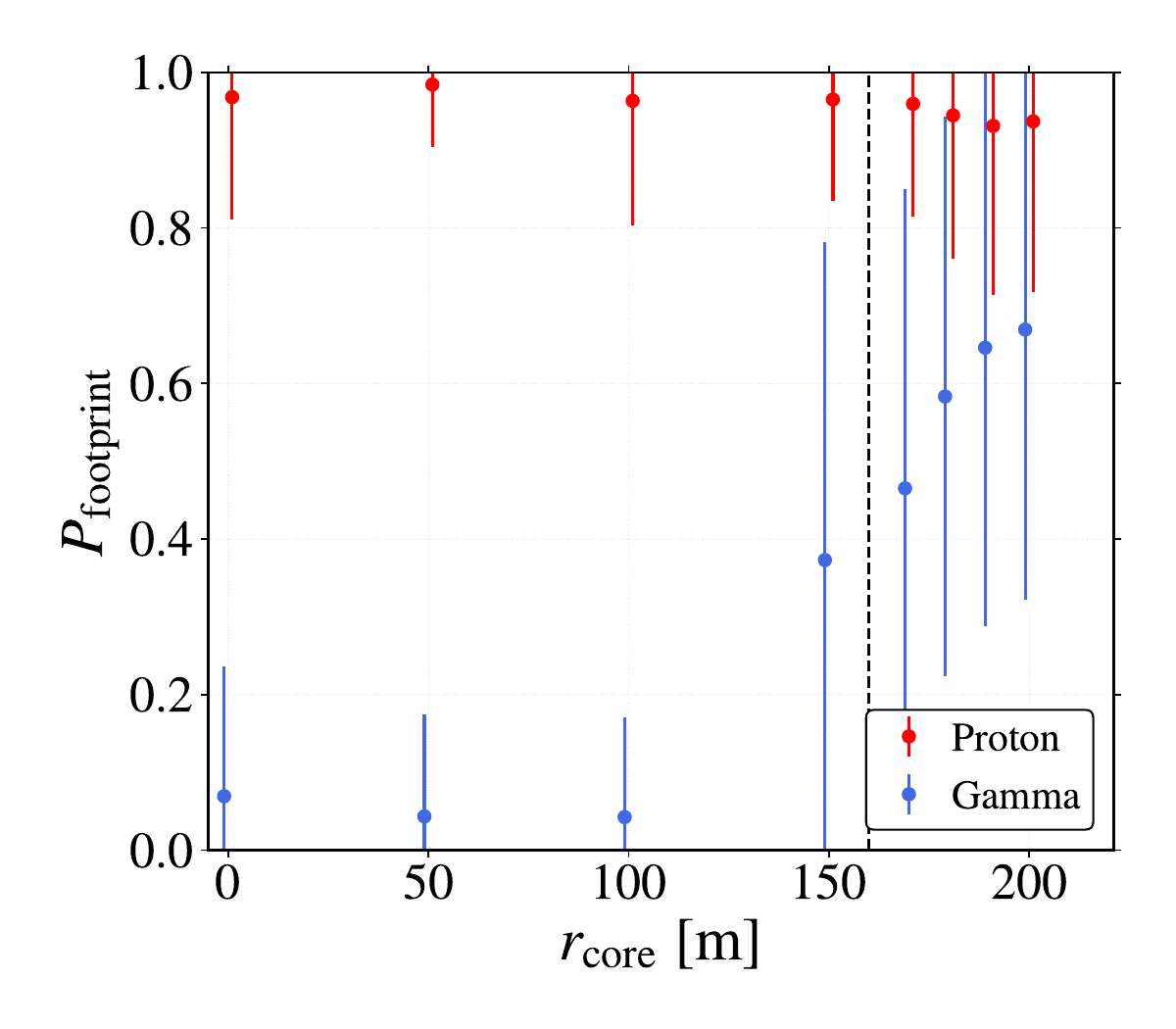}
 \caption{Average output for proton and gamma-induced showers with a reconstructed energy of roughly $500 \, \rm GeV$ in bins of the shower core distance to the center of the detector array. The dashed line indicate the limit of the array.}
 \label{fig:eff_shower_core} 
\end{figure}

The mean output probability of the model, shown in Figure~\ref{fig:eff_shower_core}, demonstrates that proton-induced showers consistently produce high probabilities close to 1, regardless of the position of the core. For gamma-induced showers, the model outputs probabilities close to 0 for on-array events, maintaining good separation from protons. However, for off-array events, the gamma probabilities increase, reducing the separation capability. Despite this, the results suggest that effective gamma/hadron discrimination remains achievable for events with core distances $r < 180\,\rm{m}$, which is beyond the array radius of $160\,\rm{m}$. It is important to note that the footprint of the shower core at these energies may cover an area of a few tens of meters (see Figures \ref{fig:example_input} and \ref{fig:example_shower_noise}), so most or all of it would be lost for off-array events. Although Figure \ref{fig:eff_shower_core} exclusively presents events in the $500 \, \rm GeV$ energy bin, similar behaviour was also observed in other energy bins.

\subsection{Dependence on the array fill factor} \label{sec:fill_factor}

\begin{figure*}[htb]
 \centering
    \subfloat[$\rm FF = 85\%$ ($5\,720$ WCDs, original).]{
   \label{fig:FF85_proton}
    \frame{\includegraphics[width=0.29\textwidth]{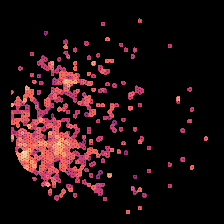}}}
    \hspace{0.05in}
    \subfloat[$\rm FF = 65\%$ ($4\,306$ WCDs).]{
 \label{fig:FF65_proton}
    \frame{\includegraphics[width=0.29\textwidth]{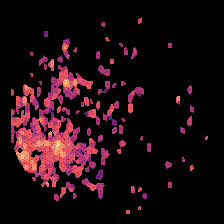}}}\hspace{0.05in}
    \subfloat[$\rm FF = 40\%$ ($2\,846$ WCDs).]{
 \label{fig:FF40_proton}
    \frame{\includegraphics[width=0.29\textwidth]{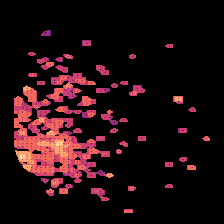}}}
 \caption{Footprint of a proton event with a reconstructed energy close to $1 \, \rm TeV$ for various array configurations exhibiting different fill factors.
}
 \label{fig:fill_factor}
\end{figure*}

To ensure that the observed performance improvements were due to the proposed method rather than external factors such as the array layout or its fill factor, two additional configurations were developed. To minimise the need for repetitive and computationally expensive Geant4 simulations, these layouts were created by uniformly reducing the number of stations in the original array (see Figure \ref{fig:fill_factor}). The original layout consisted of $5\,720$ WCDs arranged in a hexagonal pattern. The first alternative layout, with $4\,306$ WCDs and a fill factor of $65\%$, was created by removing the central station in each hexagonal grouping. The second layout featured a square arrangement of $2\,846$ WCDs, resulting in a fill factor of $40\%$. As this study focused on the sub-TeV energy range of a gamma-ray observatory, configurations with fill factors below $40\%$ were not explored.

\begin{figure}[htb]
 \centering
\includegraphics[width=0.95\linewidth]{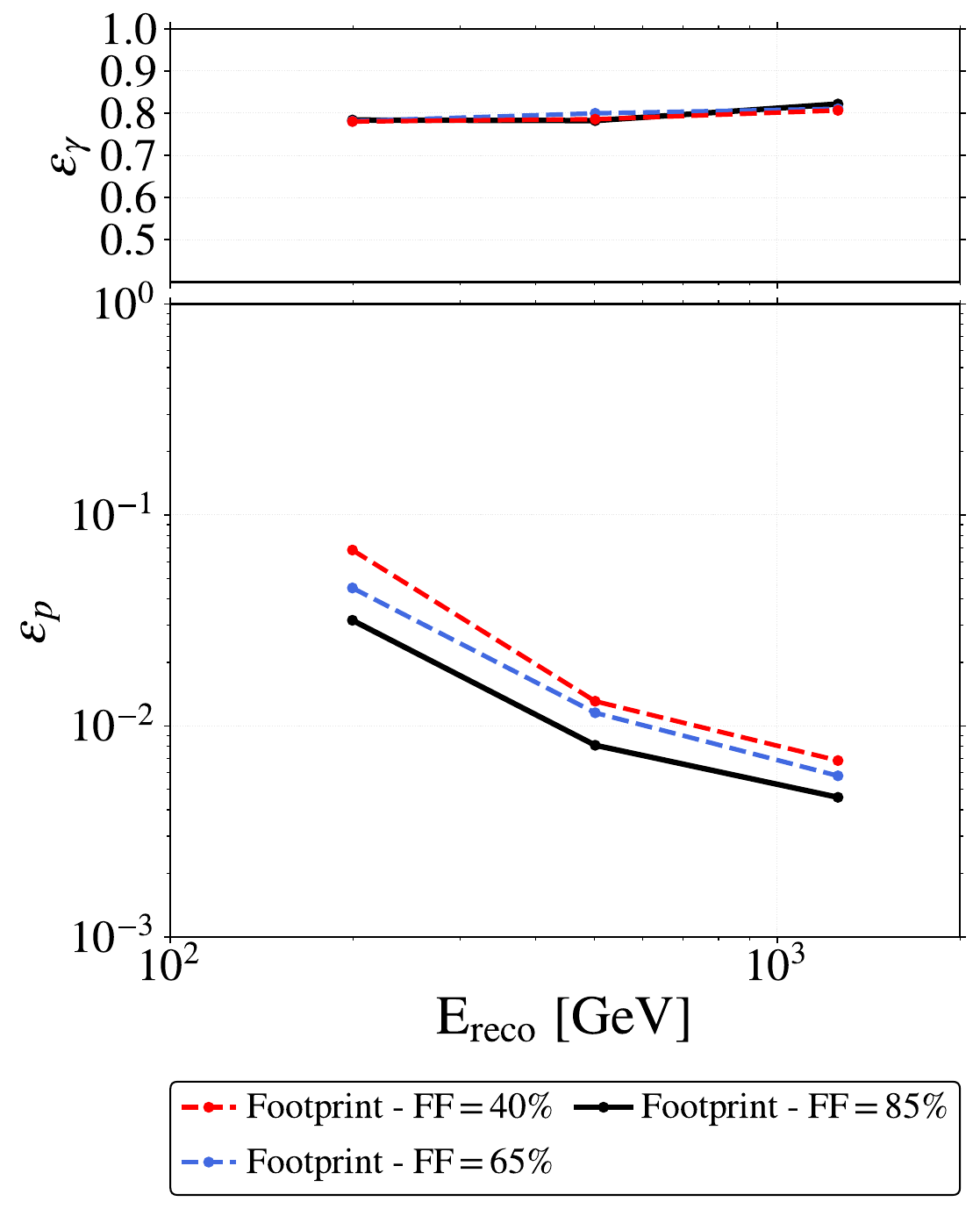}
\caption{The gamma (top) and proton (bottom) selection efficiency obtained using arrays with $\rm{ Fill \, Factor} = \left\{40;65;85 \right\}\%$.}
\label{fig:FF_comparison} 
\end{figure}

Figure \ref{fig:FF_comparison} shows the rejection efficiency achieved when using this method with the different array layouts. The models discussed in Subsection \ref{subsec:resilience} were further trained for 10 additional epochs with a subset of training showers to accommodate the new fill factors. Both configurations suffer from a performance degradation, which is more noticible at the lower energy bin. However, the model remains robust and still provides excellent gamma/hadron separation with proton rejection factors comfortably below $2 \cdot 10^{-2}$ with a high signal efficiency and $E_r \geq 500\, \rm GeV$.

\subsection{Comparison with other gamma/hadron discriminators} \label{sec:hawc_variables}

Although this method showed excellent performance compared to other observatories, this cannot be discerned from other factors such as the detector design, altitude at which the observatory is placed, or simulation setup, which can also influence the result and bias the comparison. For a fairer assessment, the gamma/hadron separators typically used by the HAWC observatory have been adapted for this specific detector design and computed for the same simulated events. Note that these discriminators originally used the PMT information due to the large size of the HAWC tanks; here, having smaller tanks, the signals at the station level were used to compute these variables. Moreover, this study did not incorporate geometry reconstruction; therefore, the Monte Carlo (MC) core position and angular measurements were employed to determine these variables. More details on the implementation of these discriminators can be found in the Appendix \ref{sec:hawc_gh_no_noise}. The following gamma/hadron separators have been implemented:

\begin{itemize}
    \item \textit{LDF} $\chi^2$: a $\chi^2$ fit of the Lateral Distribution Function (LDF) using a modified Nishimura-Kamata-Greisen (NKG) function \cite{hawc_ghsep_2024}. 
    \item Parameter for Identifying Nuclear Cosmic rays (\textit{PINCness}): a $\chi^2$-like measure of the smoothness of the charge footprint of the showers \cite{hawc_ghsep_2022}.
    \item \textit{LIC}: logarithm transformation of the inverse of the compactness parameter, an empirical parameter originally developed by the Milagro Collaboration \cite{hawc_ghsep_2022,milagro_compactness}.
\end{itemize}

\begin{figure}[htb]
 \centering
\includegraphics[width=0.95\linewidth]{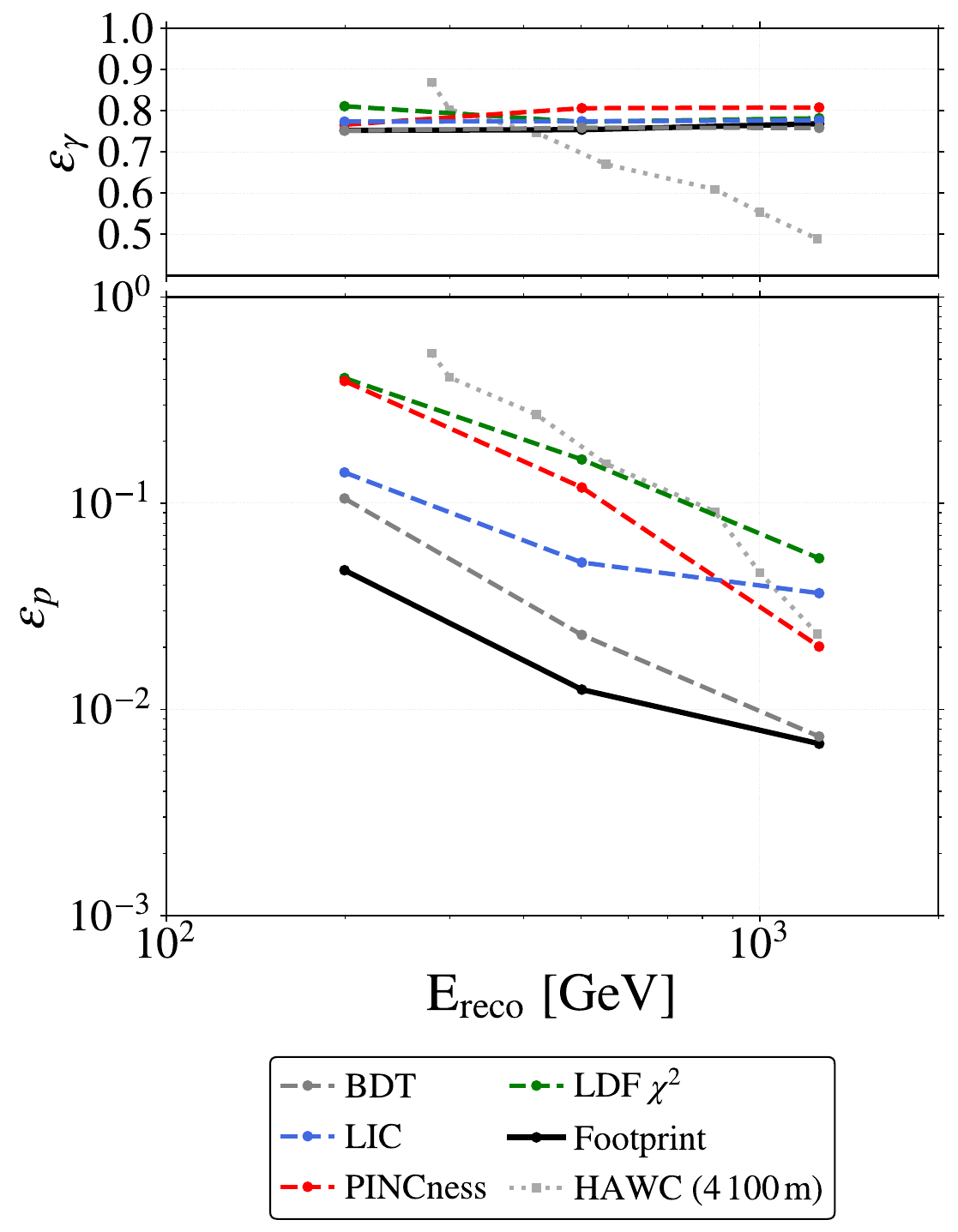}
\caption{The gamma (top) and proton (bottom) selection efficiency obtained by the HAWC gamma/hadron discriminators in showers contaminated by noise atmospheric muons and low-energy proton showers. The results integrate the shower plane fit to mitigate noise effects and are contrasted with the results reported by HAWC at a lower altitude (represented by the gray dotted line) \cite{hawc_ghsep_2024}.}
\label{fig:hawc_comparison} 
\end{figure}

\begin{figure}[htb]
 \centering
\includegraphics[width=0.8\linewidth]{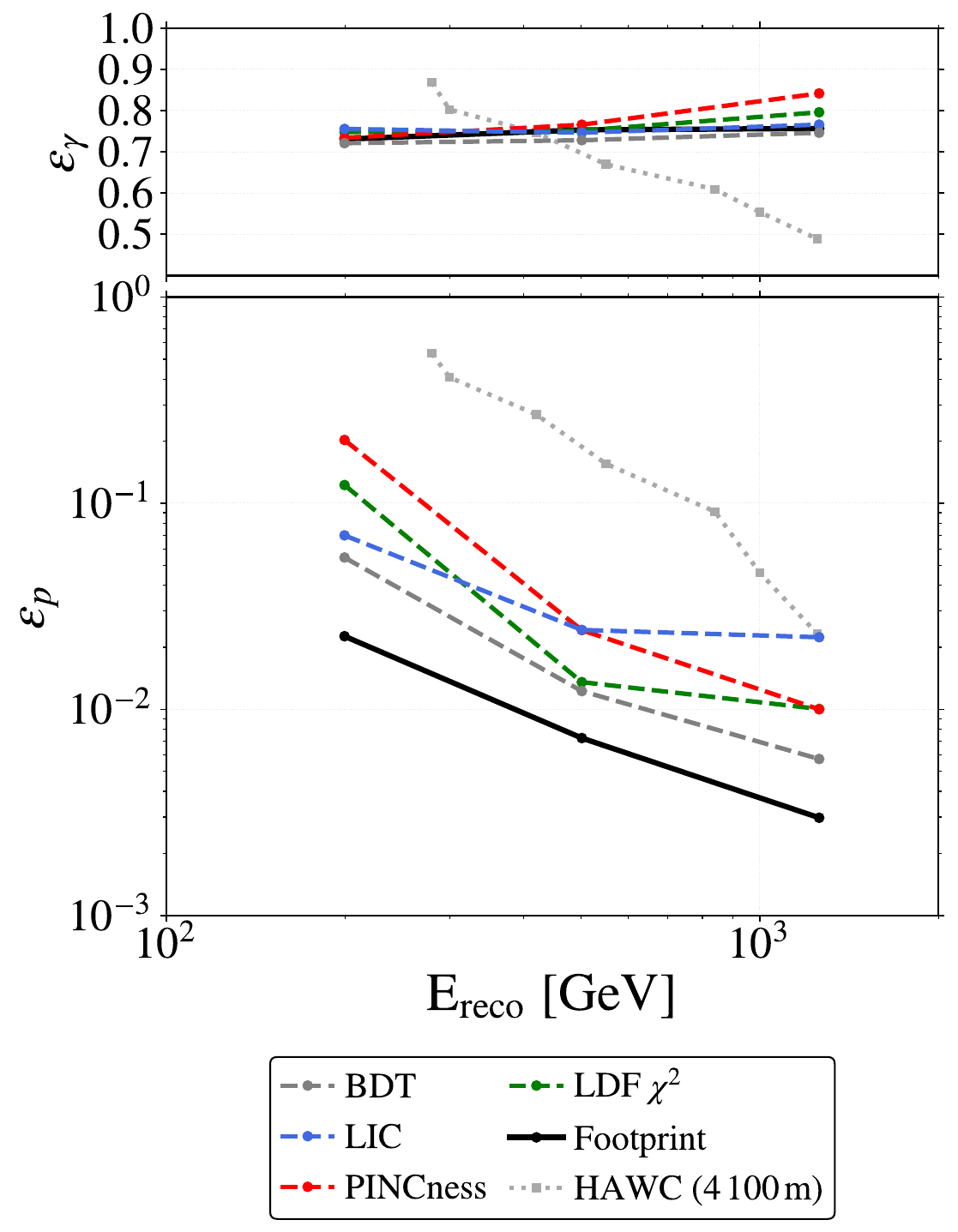}
\caption{The gamma (top) and proton (bottom) selection efficiency achieved by the HAWC gamma/hadron separators evaluated for showers without noise compared with HAWC at a lower altitude (represented by the gray dotted line) \cite{hawc_ghsep_2024}.}
\label{fig:hawc_gh_no_noise} 
\end{figure}

Figure \ref{fig:hawc_comparison} presents the proton rejection efficiency ($1-\varepsilon_p$) achieved by the gamma/hadron separators, alongside a comparison with the footprint-based method and the most recent results published by the HAWC Collaboration \cite{hawc_ghsep_2024}, which combines \textit{LDF} $\chi^2$ and \textit{compactness}. Showers with noise from both atmospheric muons and low-energy proton showers have been used to ensure a realistic comparison (the same comparison using events without noise can be found in Figure \ref{fig:hawc_gh_no_noise}). A combination of the other three discriminators that use a Boost Decision Tree (BDT) yielded significant enhancements compared to using single variables. The BDT was trained using a mean squared error loss and configured with $100$ estimators, a maximum depth of 6, and a learning rate of $0.3$. Our analysis of the alternative discriminators demonstrated improved performance relative to the results reported by HAWC. This discrepancy could be attributed to the optimistic computation of these variables using the MC core and angle values, the variation in altitude --our simulated observatory is located at $5\,200 \, \rm m$ a.s.l, whereas HAWC is situated at approximately $4\,100 \, \rm m$ a.s.l-- or the differences in the detector design.

These results also demonstrate that, under the same experimental conditions, the footprint-based method performs better than the other gamma/hadron separators, especially at the GeV energy range. It is important to note that although the plane fit helped reduce background noise, it did not completely remove it. The footprint-based method showed better resilience to noise than the other discriminators at equal conditions. For instance, the \textit{LDF} $\chi^2$ parameter showed an excellent performance for events without noise (see Figure \ref{fig:hawc_gh_no_noise}), being the best among the HAWC g/h discriminators, but noise can affect the LDF fits, causing greater degradation in classical approaches.

\subsection{Dependence on the muon content of the shower}
\begin{figure}[htb]
 \centering
\includegraphics[width=1\linewidth]{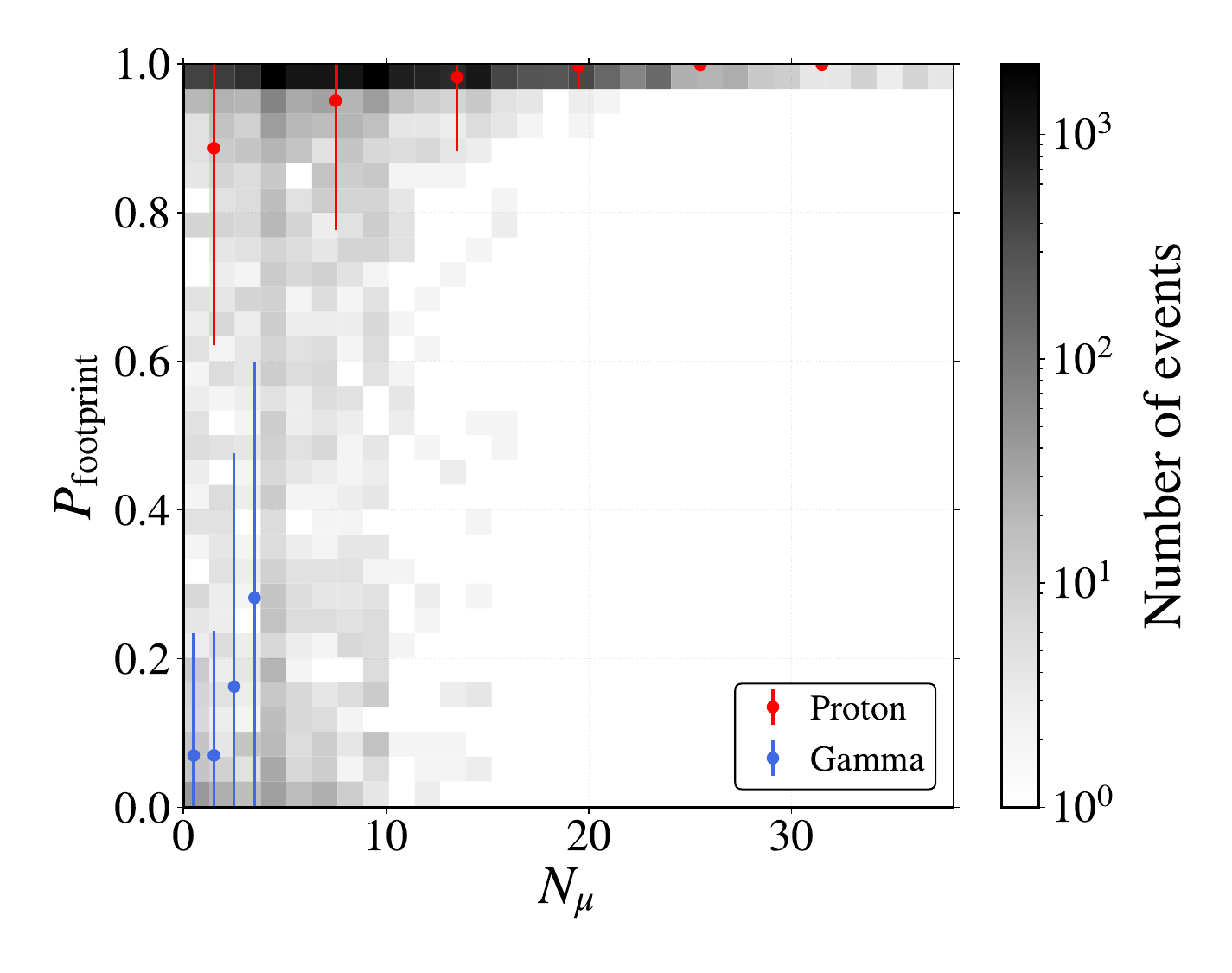}
\caption{2D histogram showing the method's probability as a function of the number of muons in stations for proton showers with $E_r \sim 500 \, \rm GeV$. It is important to note that the color bar considers only proton-induced events. Mean values are shown with error bars representing the standard deviation for both proton (red) and gamma (blue) showers in this reconstructed energy bin.}
\label{fig:eff_muon_content} 
\end{figure}

As previously noted, muon detection is one of the most effective methods for gamma/hadron separation at TeV energies, given that muons are primarily produced in hadronic interactions. To determine if the model relies on features beyond the muon content of the shower for classification, Figure \ref{fig:eff_muon_content} shows the output probability as a function of the number of stations with muons for events in the intermediate energy bin, with $E_r \sim 500 \, \rm GeV$.

The results indicate that the probability of proton-induced showers increases with the number of stations containing muons, as expected for events with a dominant hadronic component. However, significant probabilities are still observed for events with few muons, confirming that the model utilizes additional footprint features. For gamma-induced showers, a constant behaviour is observed, with most events, as expected, having no stations with muons.

\section{Conclusions}
\label{sec:conclusions}

In this study, the gamma/hadron separation capability of a footprint-based method was evaluated under various realistic conditions. Factors such as the presence of atmospheric muons, low-energy proton shower noise, variations in shower core locations, zenith angles, and array fill factors were assessed.

It was proved that, although the presence of muons enhances the performance, the method is still effective even in their absence. This demonstrates that the model is not solely dependent on muons but also leverages other footprint features for gamma/hadron separation, which is crucial at the sub-TeV energy range where muons are scarce or even absent.

The method demonstrated strong resilience to noise from low-energy proton showers and atmospheric muons, with negligible impact on the separation efficiency across the energy bins tested. Under identical conditions, this approach suggests improved performance and noise resilience compared to other state-of-the-art gamma/hadron discriminators such as the ones designed at the HAWC gamma-ray observatory.

For inclined showers with zenith angles of $\sim30^\circ$, the method proved to be effective with a similar classification performance observed for nearly vertical events. This highlights the robustness of the proposed approach, confirming its applicability to both near-vertical and inclined events.

With respect to dependence on the location of the shower core, the efficiency in discrimination for on-array events is stable, with minor variations in the average output probabilities at different core distances. However, the classification becomes more challenging for off-array events, with higher fluctuations for gamma-ray events. Nervertheless, the results suggest that the proposed method remains effective for core distances up to $180\,\rm m$, surpassing the array radius of $160\,\rm m$. This proves its robustness across various core positions, both within and near the boundaries of the array.

The investigation into different array fill factors revealed that the gamma/hadron separation efficiency is not significantly affected by the density of the detector array. Lower fill factors result in a slight reduction in classification efficiency, primarily due to the loss of granularity in the shower footprint. However, the proposed technique maintains strong separation capabilities even for a fill factor as low as $40\%$, indicating its adaptability to different detector configurations. 

Overall, this footprint-based method showed exceptional background rejection across all tested scenarios, with a performance similar to that of state-of-the-art techniques in current gamma ray experiments. These results demonstrate that there is enough information in the shower footprint to perform an outstanding gamma/hadron separation, making it a viable candidate for use in current and future gamma-ray observatories, improving their sensitivities at sub-TeV energies.


\section*{Acknowledgments}
We would like to express our gratitude to Antonio Bueno for his valuable discussions during the early stages of this work.
This work has been financed by national funds through FCT - Fundação para a Ciência e a Tecnologia, I.P., under project PTDC/FIS-PAR/4320. B.S.G. (LIP/IST) is grateful for the financial support from the FCT PhD grant PRT/BD/151553/2021 under the IDPASC program (\url{https://doi.org/10.54499/PRT/BD/151553/2021}). A.G. is grateful for the financial support from the Spanish Ministry of Economy and Competitiveness (MINECO) under the projects PID2019-104676GB-C32 and PID2023-147949NB-C53.

\bibliography{references}

\begin{thebibliography}{60}%
\makeatletter
\providecommand \@ifxundefined [1]{%
 \@ifx{#1\undefined}
}%
\providecommand \@ifnum [1]{%
 \ifnum #1\expandafter \@firstoftwo
 \else \expandafter \@secondoftwo
 \fi
}%
\providecommand \@ifx [1]{%
 \ifx #1\expandafter \@firstoftwo
 \else \expandafter \@secondoftwo
 \fi
}%
\providecommand \natexlab [1]{#1}%
\providecommand \enquote  [1]{``#1''}%
\providecommand \bibnamefont  [1]{#1}%
\providecommand \bibfnamefont [1]{#1}%
\providecommand \citenamefont [1]{#1}%
\providecommand \href@noop [0]{\@secondoftwo}%
\providecommand \href [0]{\begingroup \@sanitize@url \@href}%
\providecommand \@href[1]{\@@startlink{#1}\@@href}%
\providecommand \@@href[1]{\endgroup#1\@@endlink}%
\providecommand \@sanitize@url [0]{\catcode `\\12\catcode `\$12\catcode `\&12\catcode `\#12\catcode `\^12\catcode `\_12\catcode `\%12\relax}%
\providecommand \@@startlink[1]{}%
\providecommand \@@endlink[0]{}%
\providecommand \url  [0]{\begingroup\@sanitize@url \@url }%
\providecommand \@url [1]{\endgroup\@href {#1}{\urlprefix }}%
\providecommand \urlprefix  [0]{URL }%
\providecommand \Eprint [0]{\href }%
\providecommand \doibase [0]{https://doi.org/}%
\providecommand \selectlanguage [0]{\@gobble}%
\providecommand \bibinfo  [0]{\@secondoftwo}%
\providecommand \bibfield  [0]{\@secondoftwo}%
\providecommand \translation [1]{[#1]}%
\providecommand \BibitemOpen [0]{}%
\providecommand \bibitemStop [0]{}%
\providecommand \bibitemNoStop [0]{.\EOS\space}%
\providecommand \EOS [0]{\spacefactor3000\relax}%
\providecommand \BibitemShut  [1]{\csname bibitem#1\endcsname}%
\let\auto@bib@innerbib\@empty
\bibitem [{\citenamefont {De~Angelis}\ and\ \citenamefont {Pimenta}(2018)}]{Pimenta2018Astroparticle}%
  \BibitemOpen
  \bibfield  {author} {\bibinfo {author} {\bibfnamefont {A.}~\bibnamefont {De~Angelis}}\ and\ \bibinfo {author} {\bibfnamefont {M.}~\bibnamefont {Pimenta}},\ }\href@noop {} {\emph {\bibinfo {title} {Introduction to particle and astroparticle physics: multimessenger astronomy and its particle physics foundations}}}\ (\bibinfo  {publisher} {Springer},\ \bibinfo {year} {2018})\BibitemShut {NoStop}%
\bibitem [{\citenamefont {Viana}\ \emph {et~al.}(2019)\citenamefont {Viana}, \citenamefont {Schoorlemmer}, \citenamefont {Albert}, \citenamefont {de~Souza}, \citenamefont {Harding},\ and\ \citenamefont {Hinton}}]{viana2019searching}%
  \BibitemOpen
  \bibfield  {author} {\bibinfo {author} {\bibfnamefont {A.}~\bibnamefont {Viana}}, \bibinfo {author} {\bibfnamefont {H.}~\bibnamefont {Schoorlemmer}}, \bibinfo {author} {\bibfnamefont {A.}~\bibnamefont {Albert}}, \bibinfo {author} {\bibfnamefont {V.}~\bibnamefont {de~Souza}}, \bibinfo {author} {\bibfnamefont {J.~P.}\ \bibnamefont {Harding}},\ and\ \bibinfo {author} {\bibfnamefont {J.}~\bibnamefont {Hinton}},\ }\bibfield  {title} {\bibinfo {title} {Searching for dark matter in the galactic halo with a wide field of view tev gamma-ray observatory in the southern hemisphere},\ }\href@noop {} {\bibfield  {journal} {\bibinfo  {journal} {Journal of Cosmology and Astroparticle Physics}\ }\textbf {\bibinfo {volume} {2019}}\bibinfo  {number} { (12)},\ \bibinfo {pages} {061}}\BibitemShut {NoStop}%
\bibitem [{\citenamefont {Atwood}\ \emph {et~al.}(2009)\citenamefont {Atwood}, \citenamefont {Abdo}, \citenamefont {Ackermann}, \citenamefont {Althouse}, \citenamefont {Anderson}, \citenamefont {Axelsson}, \citenamefont {Baldini}, \citenamefont {Ballet}, \citenamefont {Band}, \citenamefont {Barbiellini} \emph {et~al.}}]{atwood2009large}%
  \BibitemOpen
\bibfield  {number} {  }\bibfield  {author} {\bibinfo {author} {\bibfnamefont {W.}~\bibnamefont {Atwood}}, \bibinfo {author} {\bibfnamefont {A.~A.}\ \bibnamefont {Abdo}}, \bibinfo {author} {\bibfnamefont {M.}~\bibnamefont {Ackermann}}, \bibinfo {author} {\bibfnamefont {W.}~\bibnamefont {Althouse}}, \bibinfo {author} {\bibfnamefont {B.}~\bibnamefont {Anderson}}, \bibinfo {author} {\bibfnamefont {M.}~\bibnamefont {Axelsson}}, \bibinfo {author} {\bibfnamefont {L.}~\bibnamefont {Baldini}}, \bibinfo {author} {\bibfnamefont {J.}~\bibnamefont {Ballet}}, \bibinfo {author} {\bibfnamefont {D.}~\bibnamefont {Band}}, \bibinfo {author} {\bibfnamefont {G.}~\bibnamefont {Barbiellini}}, \emph {et~al.},\ }\bibfield  {title} {\bibinfo {title} {The large area telescope on the fermi gamma-ray space telescope mission},\ }\href@noop {} {\bibfield  {journal} {\bibinfo  {journal} {The Astrophysical Journal}\ }\textbf {\bibinfo {volume} {697}},\ \bibinfo {pages} {1071} (\bibinfo {year} {2009})}\BibitemShut {NoStop}%
\bibitem [{\citenamefont {Degrange}\ and\ \citenamefont {Fontaine}(2015)}]{degrange2015introduction}%
  \BibitemOpen
  \bibfield  {author} {\bibinfo {author} {\bibfnamefont {B.}~\bibnamefont {Degrange}}\ and\ \bibinfo {author} {\bibfnamefont {G.}~\bibnamefont {Fontaine}},\ }\bibfield  {title} {\bibinfo {title} {Introduction to high-energy gamma-ray astronomy},\ }\href@noop {} {\bibfield  {journal} {\bibinfo  {journal} {Comptes Rendus Physique}\ }\textbf {\bibinfo {volume} {16}},\ \bibinfo {pages} {587} (\bibinfo {year} {2015})}\BibitemShut {NoStop}%
\bibitem [{\citenamefont {Albert}\ \emph {et~al.}(2019)\citenamefont {Albert}, \citenamefont {Alfaro}, \citenamefont {Ashkar}, \citenamefont {Alvarez}, \citenamefont {Alvarez}, \citenamefont {Arteaga-Vel{\'a}zquez}, \citenamefont {Solares}, \citenamefont {Arceo}, \citenamefont {Bellido}, \citenamefont {BenZvi} \emph {et~al.}}]{albert2019science}%
  \BibitemOpen
  \bibfield  {author} {\bibinfo {author} {\bibfnamefont {A.}~\bibnamefont {Albert}}, \bibinfo {author} {\bibfnamefont {R.}~\bibnamefont {Alfaro}}, \bibinfo {author} {\bibfnamefont {H.}~\bibnamefont {Ashkar}}, \bibinfo {author} {\bibfnamefont {C.}~\bibnamefont {Alvarez}}, \bibinfo {author} {\bibfnamefont {J.}~\bibnamefont {Alvarez}}, \bibinfo {author} {\bibfnamefont {J.}~\bibnamefont {Arteaga-Vel{\'a}zquez}}, \bibinfo {author} {\bibfnamefont {H.}~\bibnamefont {Solares}}, \bibinfo {author} {\bibfnamefont {R.}~\bibnamefont {Arceo}}, \bibinfo {author} {\bibfnamefont {J.}~\bibnamefont {Bellido}}, \bibinfo {author} {\bibfnamefont {S.}~\bibnamefont {BenZvi}}, \emph {et~al.},\ }\bibfield  {title} {\bibinfo {title} {Science case for a wide field-of-view very-high-energy gamma-ray observatory in the southern hemisphere},\ }\href@noop {} {\bibfield  {journal} {\bibinfo  {journal} {arXiv preprint arXiv:1902.08429}\ } (\bibinfo {year} {2019})}\BibitemShut {NoStop}%
\bibitem [{\citenamefont {Torres~Peralta}\ \emph {et~al.}(2024)\citenamefont {Torres~Peralta} \emph {et~al.}}]{torres2024enhanced}%
  \BibitemOpen
  \bibfield  {author} {\bibinfo {author} {\bibfnamefont {T.~J.}\ \bibnamefont {Torres~Peralta}} \emph {et~al.} (\bibinfo {collaboration} {LAGO}),\ }\bibfield  {title} {\bibinfo {title} {Enhanced particle classification in water cherenkov detectors using machine learning: Modeling and validation with monte carlo simulation datasets},\ }\href@noop {} {\bibfield  {journal} {\bibinfo  {journal} {Atmosphere}\ }\textbf {\bibinfo {volume} {15}},\ \bibinfo {pages} {1039} (\bibinfo {year} {2024})}\BibitemShut {NoStop}%
\bibitem [{\citenamefont {Assis}\ \emph {et~al.}(2022)\citenamefont {Assis}, \citenamefont {Bakalov{\'a}}, \citenamefont {Barres~de Almeida}, \citenamefont {Brogueira}, \citenamefont {Concei{\c{c}}{\~a}o}, \citenamefont {De~Angelis}, \citenamefont {Gibilisco}, \citenamefont {Gonz{\'a}lez}, \citenamefont {Guill{\'e}n}, \citenamefont {La~Mura} \emph {et~al.}}]{wcd2022mercedes}%
  \BibitemOpen
  \bibfield  {author} {\bibinfo {author} {\bibfnamefont {P.}~\bibnamefont {Assis}}, \bibinfo {author} {\bibfnamefont {A.}~\bibnamefont {Bakalov{\'a}}}, \bibinfo {author} {\bibfnamefont {U.}~\bibnamefont {Barres~de Almeida}}, \bibinfo {author} {\bibfnamefont {P.}~\bibnamefont {Brogueira}}, \bibinfo {author} {\bibfnamefont {R.}~\bibnamefont {Concei{\c{c}}{\~a}o}}, \bibinfo {author} {\bibfnamefont {A.}~\bibnamefont {De~Angelis}}, \bibinfo {author} {\bibfnamefont {L.}~\bibnamefont {Gibilisco}}, \bibinfo {author} {\bibfnamefont {B.}~\bibnamefont {Gonz{\'a}lez}}, \bibinfo {author} {\bibfnamefont {A.}~\bibnamefont {Guill{\'e}n}}, \bibinfo {author} {\bibfnamefont {G.}~\bibnamefont {La~Mura}}, \emph {et~al.},\ }\bibfield  {title} {\bibinfo {title} {The mercedes water cherenkov detector},\ }\href@noop {} {\bibfield  {journal} {\bibinfo  {journal} {The European Physical Journal C}\ }\textbf {\bibinfo {volume} {82}},\ \bibinfo {pages} {899} (\bibinfo {year} {2022})}\BibitemShut {NoStop}%
\bibitem [{\citenamefont {Concei\c{c}\~ao}\ \emph {et~al.}(2021)\citenamefont {Concei\c{c}\~ao}, \citenamefont {Gonz\'alez}, \citenamefont {Guill\'en}, \citenamefont {Pimenta},\ and\ \citenamefont {Tom\'e}}]{WCD4PMTs}%
  \BibitemOpen
  \bibfield  {author} {\bibinfo {author} {\bibfnamefont {R.}~\bibnamefont {Concei\c{c}\~ao}}, \bibinfo {author} {\bibfnamefont {B.~S.}\ \bibnamefont {Gonz\'alez}}, \bibinfo {author} {\bibfnamefont {A.}~\bibnamefont {Guill\'en}}, \bibinfo {author} {\bibfnamefont {M.}~\bibnamefont {Pimenta}},\ and\ \bibinfo {author} {\bibfnamefont {B.}~\bibnamefont {Tom\'e}},\ }\bibfield  {title} {\bibinfo {title} {{Muon identification in a compact single-layered water Cherenkov detector and gamma/hadron discrimination using machine learning techniques}},\ }\href {https://doi.org/10.1140/epjc/s10052-021-09312-4} {\bibfield  {journal} {\bibinfo  {journal} {Eur. Phys. J. C}\ }\textbf {\bibinfo {volume} {81}},\ \bibinfo {pages} {542} (\bibinfo {year} {2021})},\ \Eprint {https://arxiv.org/abs/2101.10109} {arXiv:2101.10109 [physics.ins-det]} \BibitemShut {NoStop}%
\bibitem [{\citenamefont {Gonz{\'a}lez}\ \emph {et~al.}(2022)\citenamefont {Gonz{\'a}lez}, \citenamefont {Concei{\c{c}}{\~a}o}, \citenamefont {Pimenta}, \citenamefont {Tom{\'e}},\ and\ \citenamefont {Guill{\'e}n}}]{4PMTs_NCA}%
  \BibitemOpen
  \bibfield  {author} {\bibinfo {author} {\bibfnamefont {B.~S.}\ \bibnamefont {Gonz{\'a}lez}}, \bibinfo {author} {\bibfnamefont {R.}~\bibnamefont {Concei{\c{c}}{\~a}o}}, \bibinfo {author} {\bibfnamefont {M.}~\bibnamefont {Pimenta}}, \bibinfo {author} {\bibfnamefont {B.}~\bibnamefont {Tom{\'e}}},\ and\ \bibinfo {author} {\bibfnamefont {A.}~\bibnamefont {Guill{\'e}n}},\ }\bibfield  {title} {\bibinfo {title} {Tackling the muon identification in water cherenkov detectors problem for the future southern wide-field gamma-ray observatory by means of machine learning},\ }\href {https://doi.org/https://doi.org/10.1007/s00521-021-06730-z} {\bibfield  {journal} {\bibinfo  {journal} {Neural Computing and Applications}\ ,\ \bibinfo {pages} {1}} (\bibinfo {year} {2022})},\ \Eprint {https://arxiv.org/abs/2101.11924} {arXiv:2101.11924 [physics.ins-det]} \BibitemShut {NoStop}%
\bibitem [{\citenamefont {González}\ \emph {et~al.}(2020)\citenamefont {González}, \citenamefont {Conceição}, \citenamefont {Tomé}, \citenamefont {Pimenta}, \citenamefont {Herrera},\ and\ \citenamefont {Guillen}}]{WCD_9iPM_2020}%
  \BibitemOpen
  \bibfield  {author} {\bibinfo {author} {\bibfnamefont {B.}~\bibnamefont {González}}, \bibinfo {author} {\bibfnamefont {R.}~\bibnamefont {Conceição}}, \bibinfo {author} {\bibfnamefont {B.}~\bibnamefont {Tomé}}, \bibinfo {author} {\bibfnamefont {M.}~\bibnamefont {Pimenta}}, \bibinfo {author} {\bibfnamefont {L.}~\bibnamefont {Herrera}},\ and\ \bibinfo {author} {\bibfnamefont {A.}~\bibnamefont {Guillen}},\ }\bibfield  {title} {\bibinfo {title} {Using convolutional neural networks for muon detection in wcd tank},\ }\href {https://doi.org/10.1088/1742-6596/1603/1/012024} {\bibfield  {journal} {\bibinfo  {journal} {Journal of Physics: Conference Series}\ }\textbf {\bibinfo {volume} {1603}},\ \bibinfo {pages} {012024} (\bibinfo {year} {2020})}\BibitemShut {NoStop}%
\bibitem [{\citenamefont {Kunwar}\ \emph {et~al.}(2023)\citenamefont {Kunwar}, \citenamefont {Goksu}, \citenamefont {Hinton}, \citenamefont {Schoorlemmer}, \citenamefont {Smith}, \citenamefont {Hofmann},\ and\ \citenamefont {Werner}}]{kunwar2023double}%
  \BibitemOpen
  \bibfield  {author} {\bibinfo {author} {\bibfnamefont {S.}~\bibnamefont {Kunwar}}, \bibinfo {author} {\bibfnamefont {H.}~\bibnamefont {Goksu}}, \bibinfo {author} {\bibfnamefont {J.}~\bibnamefont {Hinton}}, \bibinfo {author} {\bibfnamefont {H.}~\bibnamefont {Schoorlemmer}}, \bibinfo {author} {\bibfnamefont {A.}~\bibnamefont {Smith}}, \bibinfo {author} {\bibfnamefont {W.}~\bibnamefont {Hofmann}},\ and\ \bibinfo {author} {\bibfnamefont {F.}~\bibnamefont {Werner}},\ }\bibfield  {title} {\bibinfo {title} {A double-layered water cherenkov detector array for gamma-ray astronomy},\ }\href@noop {} {\bibfield  {journal} {\bibinfo  {journal} {Nuclear Instruments and Methods in Physics Research Section A: Accelerators, Spectrometers, Detectors and Associated Equipment}\ }\textbf {\bibinfo {volume} {1050}},\ \bibinfo {pages} {168138} (\bibinfo {year} {2023})}\BibitemShut {NoStop}%
\bibitem [{\citenamefont {Iacovacci}\ \emph {et~al.}(2013)\citenamefont {Iacovacci}, \citenamefont {{Di Girolamo}}, \citenamefont {Mastroianni},\ and\ \citenamefont {Li}}]{ARGO_2013}%
  \BibitemOpen
  \bibfield  {author} {\bibinfo {author} {\bibfnamefont {M.}~\bibnamefont {Iacovacci}}, \bibinfo {author} {\bibfnamefont {T.}~\bibnamefont {{Di Girolamo}}}, \bibinfo {author} {\bibfnamefont {S.}~\bibnamefont {Mastroianni}},\ and\ \bibinfo {author} {\bibfnamefont {X.}~\bibnamefont {Li}},\ }\bibfield  {title} {\bibinfo {title} {Spatial correlations applied to gamma/hadron discrimination in the argo-ybj experiment},\ }\href {https://doi.org/https://doi.org/10.1016/j.nuclphysbps.2013.05.038} {\bibfield  {journal} {\bibinfo  {journal} {Nuclear Physics B - Proceedings Supplements}\ }\textbf {\bibinfo {volume} {239-240}},\ \bibinfo {pages} {250} (\bibinfo {year} {2013})},\ \bibinfo {note} {proceedings of the 9th workshop on Science with the New Generation of High Energy Gamma-ray Experiments: From high energy gamma sources to cosmic rays, one century after their discovery}\BibitemShut {NoStop}%
\bibitem [{\citenamefont {Concei\c{c}\~ao}\ \emph {et~al.}(2022)\citenamefont {Concei\c{c}\~ao}, \citenamefont {Gibilisco}, \citenamefont {Pimenta},\ and\ \citenamefont {Tom{\'e}}}]{LCm2022gamma}%
  \BibitemOpen
  \bibfield  {author} {\bibinfo {author} {\bibfnamefont {R.}~\bibnamefont {Concei\c{c}\~ao}}, \bibinfo {author} {\bibfnamefont {L.}~\bibnamefont {Gibilisco}}, \bibinfo {author} {\bibfnamefont {M.}~\bibnamefont {Pimenta}},\ and\ \bibinfo {author} {\bibfnamefont {B.}~\bibnamefont {Tom{\'e}}},\ }\bibfield  {title} {\bibinfo {title} {Gamma/hadron discrimination at high energies through the azimuthal fluctuations of air shower particle distributions at the ground},\ }\href@noop {} {\bibfield  {journal} {\bibinfo  {journal} {Journal of Cosmology and Astroparticle Physics}\ }\textbf {\bibinfo {volume} {2022}}\bibinfo  {number} { (10)},\ \bibinfo {pages} {086}}\BibitemShut {NoStop}%
\bibitem [{\citenamefont {Concei\c{c}\~ao}\ \emph {et~al.}(2024)\citenamefont {Concei\c{c}\~ao}, \citenamefont {Costa}, \citenamefont {Gibilisco}, \citenamefont {Pimenta},\ and\ \citenamefont {Tom\'e}}]{Ptail2024}%
  \BibitemOpen
\bibfield  {number} {  }\bibfield  {author} {\bibinfo {author} {\bibfnamefont {R.}~\bibnamefont {Concei\c{c}\~ao}}, \bibinfo {author} {\bibfnamefont {P.~J.}\ \bibnamefont {Costa}}, \bibinfo {author} {\bibfnamefont {L.}~\bibnamefont {Gibilisco}}, \bibinfo {author} {\bibfnamefont {M.}~\bibnamefont {Pimenta}},\ and\ \bibinfo {author} {\bibfnamefont {B.}~\bibnamefont {Tom\'e}},\ }\bibfield  {title} {\bibinfo {title} {High resolution gamma/hadron and composition discriminant variable for water-cherenkov detector cosmic-ray observatories},\ }\href {https://doi.org/10.1103/PhysRevD.110.023033} {\bibfield  {journal} {\bibinfo  {journal} {Phys. Rev. D}\ }\textbf {\bibinfo {volume} {110}},\ \bibinfo {pages} {023033} (\bibinfo {year} {2024})}\BibitemShut {NoStop}%
\bibitem [{\citenamefont {Abeysekara}\ \emph {et~al.}(2017)\citenamefont {Abeysekara} \emph {et~al.}}]{HAWC_Crab}%
  \BibitemOpen
  \bibfield  {author} {\bibinfo {author} {\bibfnamefont {A.}~\bibnamefont {Abeysekara}} \emph {et~al.},\ }\bibfield  {title} {\bibinfo {title} {{Daily monitoring of TeV gamma-ray emission from Mrk 421, Mrk 501, and the Crab Nebula with HAWC}},\ }\href {https://doi.org/10.3847/1538-4357/aa729e} {\bibfield  {journal} {\bibinfo  {journal} {Astrophys. J.}\ }\textbf {\bibinfo {volume} {841}},\ \bibinfo {pages} {100} (\bibinfo {year} {2017})},\ \Eprint {https://arxiv.org/abs/1703.06968} {arXiv:1703.06968 [astro-ph.HE]} \BibitemShut {NoStop}%
\bibitem [{\citenamefont {Assun{\c{c}}{\~a}o}\ \emph {et~al.}(2019)\citenamefont {Assun{\c{c}}{\~a}o}, \citenamefont {Correia}, \citenamefont {Concei{\c{c}}{\~a}o}, \citenamefont {Pimenta}, \citenamefont {Tom{\'e}}, \citenamefont {Louren{\c{c}}o},\ and\ \citenamefont {Machado}}]{assunccao2019automatic}%
  \BibitemOpen
  \bibfield  {author} {\bibinfo {author} {\bibfnamefont {F.}~\bibnamefont {Assun{\c{c}}{\~a}o}}, \bibinfo {author} {\bibfnamefont {J.}~\bibnamefont {Correia}}, \bibinfo {author} {\bibfnamefont {R.}~\bibnamefont {Concei{\c{c}}{\~a}o}}, \bibinfo {author} {\bibfnamefont {M.~J.~M.}\ \bibnamefont {Pimenta}}, \bibinfo {author} {\bibfnamefont {B.}~\bibnamefont {Tom{\'e}}}, \bibinfo {author} {\bibfnamefont {N.}~\bibnamefont {Louren{\c{c}}o}},\ and\ \bibinfo {author} {\bibfnamefont {P.}~\bibnamefont {Machado}},\ }\bibfield  {title} {\bibinfo {title} {Automatic design of artificial neural networks for gamma-ray detection},\ }\href@noop {} {\bibfield  {journal} {\bibinfo  {journal} {IEEE Access}\ }\textbf {\bibinfo {volume} {7}},\ \bibinfo {pages} {110531} (\bibinfo {year} {2019})}\BibitemShut {NoStop}%
\bibitem [{\citenamefont {Glombitza}\ \emph {et~al.}(2024)\citenamefont {Glombitza}, \citenamefont {Schneider}, \citenamefont {Leitl}, \citenamefont {Funk},\ and\ \citenamefont {van Eldik}}]{gnn2024jonas}%
  \BibitemOpen
  \bibfield  {author} {\bibinfo {author} {\bibfnamefont {J.}~\bibnamefont {Glombitza}}, \bibinfo {author} {\bibfnamefont {M.}~\bibnamefont {Schneider}}, \bibinfo {author} {\bibfnamefont {F.}~\bibnamefont {Leitl}}, \bibinfo {author} {\bibfnamefont {S.}~\bibnamefont {Funk}},\ and\ \bibinfo {author} {\bibfnamefont {C.}~\bibnamefont {van Eldik}},\ }\bibfield  {title} {\bibinfo {title} {Application of graph networks to a wide-field water-cherenkov-based gamma-ray observatory},\ }\href@noop {} {\bibfield  {journal} {\bibinfo  {journal} {arXiv preprint arXiv:2411.16565}\ } (\bibinfo {year} {2024})}\BibitemShut {NoStop}%
\bibitem [{\citenamefont {Shilon}\ \emph {et~al.}(2019)\citenamefont {Shilon}, \citenamefont {Kraus}, \citenamefont {B{\"u}chele}, \citenamefont {Egberts}, \citenamefont {Fischer}, \citenamefont {Holch}, \citenamefont {Lohse}, \citenamefont {Schwanke}, \citenamefont {Steppa},\ and\ \citenamefont {Funk}}]{Shilon2019}%
  \BibitemOpen
  \bibfield  {author} {\bibinfo {author} {\bibfnamefont {I.}~\bibnamefont {Shilon}}, \bibinfo {author} {\bibfnamefont {M.}~\bibnamefont {Kraus}}, \bibinfo {author} {\bibfnamefont {M.}~\bibnamefont {B{\"u}chele}}, \bibinfo {author} {\bibfnamefont {K.}~\bibnamefont {Egberts}}, \bibinfo {author} {\bibfnamefont {T.}~\bibnamefont {Fischer}}, \bibinfo {author} {\bibfnamefont {T.~L.}\ \bibnamefont {Holch}}, \bibinfo {author} {\bibfnamefont {T.}~\bibnamefont {Lohse}}, \bibinfo {author} {\bibfnamefont {U.}~\bibnamefont {Schwanke}}, \bibinfo {author} {\bibfnamefont {C.}~\bibnamefont {Steppa}},\ and\ \bibinfo {author} {\bibfnamefont {S.}~\bibnamefont {Funk}},\ }\bibfield  {title} {\bibinfo {title} {Application of deep learning methods to analysis of imaging atmospheric cherenkov telescopes data},\ }\href@noop {} {\bibfield  {journal} {\bibinfo  {journal} {Astroparticle Physics}\ }\textbf {\bibinfo {volume} {105}},\ \bibinfo {pages} {44} (\bibinfo {year} {2019})}\BibitemShut {NoStop}%
\bibitem [{\citenamefont {Glombitza}\ \emph {et~al.}(2023)\citenamefont {Glombitza}, \citenamefont {Joshi}, \citenamefont {Bruno},\ and\ \citenamefont {Funk}}]{jonas_gnn_IACT_2023}%
  \BibitemOpen
  \bibfield  {author} {\bibinfo {author} {\bibfnamefont {J.}~\bibnamefont {Glombitza}}, \bibinfo {author} {\bibfnamefont {V.}~\bibnamefont {Joshi}}, \bibinfo {author} {\bibfnamefont {B.}~\bibnamefont {Bruno}},\ and\ \bibinfo {author} {\bibfnamefont {S.}~\bibnamefont {Funk}},\ }\bibfield  {title} {\bibinfo {title} {Application of graph networks to background rejection in imaging air cherenkov telescopes},\ }\href@noop {} {\bibfield  {journal} {\bibinfo  {journal} {Journal of Cosmology and Astroparticle Physics}\ }\textbf {\bibinfo {volume} {2023}}\bibinfo  {number} { (11)},\ \bibinfo {pages} {008}}\BibitemShut {NoStop}%
\bibitem [{\citenamefont {Okukawa}\ \emph {et~al.}(2024)\citenamefont {Okukawa}, \citenamefont {Hara}, \citenamefont {Hibino}, \citenamefont {Katayose}, \citenamefont {Kawata}, \citenamefont {Ohnishi}, \citenamefont {Sako}, \citenamefont {Sako}, \citenamefont {Shibata}, \citenamefont {Shiomi} \emph {et~al.}}]{ghsep_2024_tibet}%
  \BibitemOpen
\bibfield  {number} {  }\bibfield  {author} {\bibinfo {author} {\bibfnamefont {S.}~\bibnamefont {Okukawa}}, \bibinfo {author} {\bibfnamefont {K.}~\bibnamefont {Hara}}, \bibinfo {author} {\bibfnamefont {K.}~\bibnamefont {Hibino}}, \bibinfo {author} {\bibfnamefont {Y.}~\bibnamefont {Katayose}}, \bibinfo {author} {\bibfnamefont {K.}~\bibnamefont {Kawata}}, \bibinfo {author} {\bibfnamefont {M.}~\bibnamefont {Ohnishi}}, \bibinfo {author} {\bibfnamefont {T.}~\bibnamefont {Sako}}, \bibinfo {author} {\bibfnamefont {T.~K.}\ \bibnamefont {Sako}}, \bibinfo {author} {\bibfnamefont {M.}~\bibnamefont {Shibata}}, \bibinfo {author} {\bibfnamefont {A.}~\bibnamefont {Shiomi}}, \emph {et~al.},\ }\bibfield  {title} {\bibinfo {title} {Neural networks for separation of cosmic gamma rays and hadronic cosmic rays in air shower observation with a large area surface detector array},\ }\href@noop {} {\bibfield  {journal} {\bibinfo  {journal} {Machine Learning: Science and Technology}\ }\textbf {\bibinfo {volume} {5}},\ \bibinfo {pages}
  {025016} (\bibinfo {year} {2024})}\BibitemShut {NoStop}%
\bibitem [{\citenamefont {Wang}\ \emph {et~al.}(2019)\citenamefont {Wang} \emph {et~al.}}]{lhaaso_ghsep_2020}%
  \BibitemOpen
  \bibfield  {author} {\bibinfo {author} {\bibfnamefont {X.}~\bibnamefont {Wang}} \emph {et~al.} (\bibinfo {collaboration} {LHAASO}),\ }\bibfield  {title} {\bibinfo {title} {{Gamma Hadron separation using traditional single parameter method and multivariate algorithms with LHAASO-WCDA experiment}},\ }\href {https://doi.org/10.22323/1.358.0820} {\bibfield  {journal} {\bibinfo  {journal} {PoS}\ }\textbf {\bibinfo {volume} {ICRC2019}},\ \bibinfo {pages} {820} (\bibinfo {year} {2019})}\BibitemShut {NoStop}%
\bibitem [{\citenamefont {Aharonian}\ \emph {et~al.}(2021)\citenamefont {Aharonian} \emph {et~al.}}]{LHAASO2021performance}%
  \BibitemOpen
  \bibfield  {author} {\bibinfo {author} {\bibfnamefont {F.}~\bibnamefont {Aharonian}} \emph {et~al.} (\bibinfo {collaboration} {LHAASO}),\ }\bibfield  {title} {\bibinfo {title} {Performance of lhaaso-wcda and observation of the crab nebula as a standard candle},\ }\href@noop {} {\bibfield  {journal} {\bibinfo  {journal} {Chinese Physics C}\ }\textbf {\bibinfo {volume} {45}},\ \bibinfo {pages} {085002} (\bibinfo {year} {2021})}\BibitemShut {NoStop}%
\bibitem [{\citenamefont {Jin}\ \emph {et~al.}(2020)\citenamefont {Jin}, \citenamefont {Chen}, \citenamefont {He} \emph {et~al.}}]{LHAASO_2020_gnn}%
  \BibitemOpen
  \bibfield  {author} {\bibinfo {author} {\bibfnamefont {C.}~\bibnamefont {Jin}}, \bibinfo {author} {\bibfnamefont {S.-z.}\ \bibnamefont {Chen}}, \bibinfo {author} {\bibfnamefont {H.-h.}\ \bibnamefont {He}}, \emph {et~al.} (\bibinfo {collaboration} {LHAASO}),\ }\bibfield  {title} {\bibinfo {title} {Classifying cosmic-ray proton and light groups in lhaaso-km2a experiment with graph neural network},\ }\href@noop {} {\bibfield  {journal} {\bibinfo  {journal} {Chinese Physics C}\ }\textbf {\bibinfo {volume} {44}},\ \bibinfo {pages} {065002} (\bibinfo {year} {2020})}\BibitemShut {NoStop}%
\bibitem [{\citenamefont {Capistr\'an}\ \emph {et~al.}(2021)\citenamefont {Capistr\'an} \emph {et~al.}}]{HAWC2021ML}%
  \BibitemOpen
  \bibfield  {author} {\bibinfo {author} {\bibfnamefont {T.}~\bibnamefont {Capistr\'an}} \emph {et~al.} (\bibinfo {collaboration} {HAWC}),\ }\bibfield  {title} {\bibinfo {title} {{Use of Machine Learning for gamma/hadron separation with HAWC}},\ }\href {https://doi.org/10.22323/1.395.0745} {\bibfield  {journal} {\bibinfo  {journal} {PoS}\ }\textbf {\bibinfo {volume} {ICRC2021}},\ \bibinfo {pages} {745} (\bibinfo {year} {2021})},\ \Eprint {https://arxiv.org/abs/2108.00112} {arXiv:2108.00112 [astro-ph.HE]} \BibitemShut {NoStop}%
\bibitem [{\citenamefont {Carrillo-Perez}\ \emph {et~al.}(2019)\citenamefont {Carrillo-Perez}, \citenamefont {Herrera}, \citenamefont {Carceller},\ and\ \citenamefont {Guill{\'e}n}}]{carrillo2019improving}%
  \BibitemOpen
  \bibfield  {author} {\bibinfo {author} {\bibfnamefont {F.}~\bibnamefont {Carrillo-Perez}}, \bibinfo {author} {\bibfnamefont {L.~J.}\ \bibnamefont {Herrera}}, \bibinfo {author} {\bibfnamefont {J.~M.}\ \bibnamefont {Carceller}},\ and\ \bibinfo {author} {\bibfnamefont {A.}~\bibnamefont {Guill{\'e}n}},\ }\bibfield  {title} {\bibinfo {title} {Improving classification of ultra-high energy cosmic rays using spacial locality by means of a convolutional dnn},\ }in\ \href@noop {} {\emph {\bibinfo {booktitle} {International Work-Conference on Artificial Neural Networks}}}\ (\bibinfo {organization} {Springer},\ \bibinfo {year} {2019})\ pp.\ \bibinfo {pages} {222--232}\BibitemShut {NoStop}%
\bibitem [{\citenamefont {Guill{\'e}n}\ \emph {et~al.}(2019)\citenamefont {Guill{\'e}n}, \citenamefont {Bueno}, \citenamefont {Carceller}, \citenamefont {Mart{\'i}nez-Vel{\'a}zquez}, \citenamefont {Rubio}, \citenamefont {Peixoto},\ and\ \citenamefont {Sanchez-Lucas}}]{GUILLEN201912}%
  \BibitemOpen
  \bibfield  {author} {\bibinfo {author} {\bibfnamefont {A.}~\bibnamefont {Guill{\'e}n}}, \bibinfo {author} {\bibfnamefont {A.}~\bibnamefont {Bueno}}, \bibinfo {author} {\bibfnamefont {J.}~\bibnamefont {Carceller}}, \bibinfo {author} {\bibfnamefont {J.}~\bibnamefont {Mart{\'i}nez-Vel{\'a}zquez}}, \bibinfo {author} {\bibfnamefont {G.}~\bibnamefont {Rubio}}, \bibinfo {author} {\bibfnamefont {C.~T.}\ \bibnamefont {Peixoto}},\ and\ \bibinfo {author} {\bibfnamefont {P.}~\bibnamefont {Sanchez-Lucas}},\ }\bibfield  {title} {\bibinfo {title} {Deep learning techniques applied to the physics of extensive air showers},\ }\href {https://doi.org/https://doi.org/10.1016/j.astropartphys.2019.03.001} {\bibfield  {journal} {\bibinfo  {journal} {Astroparticle Physics}\ }\textbf {\bibinfo {volume} {111}},\ \bibinfo {pages} {12 } (\bibinfo {year} {2019})}\BibitemShut {NoStop}%
\bibitem [{\citenamefont {Nieto}\ \emph {et~al.}(2019)\citenamefont {Nieto}, \citenamefont {Brill}, \citenamefont {Feng}, \citenamefont {Humensky}, \citenamefont {Kim}, \citenamefont {Miener}, \citenamefont {Mukherjee},\ and\ \citenamefont {Sevilla}}]{nieto2019ctlearn}%
  \BibitemOpen
  \bibfield  {author} {\bibinfo {author} {\bibfnamefont {D.}~\bibnamefont {Nieto}}, \bibinfo {author} {\bibfnamefont {A.}~\bibnamefont {Brill}}, \bibinfo {author} {\bibfnamefont {Q.}~\bibnamefont {Feng}}, \bibinfo {author} {\bibfnamefont {T.}~\bibnamefont {Humensky}}, \bibinfo {author} {\bibfnamefont {B.}~\bibnamefont {Kim}}, \bibinfo {author} {\bibfnamefont {T.}~\bibnamefont {Miener}}, \bibinfo {author} {\bibfnamefont {R.}~\bibnamefont {Mukherjee}},\ and\ \bibinfo {author} {\bibfnamefont {J.}~\bibnamefont {Sevilla}},\ }\bibfield  {title} {\bibinfo {title} {Ctlearn: Deep learning for gamma-ray astronomy},\ }\href@noop {} {\bibfield  {journal} {\bibinfo  {journal} {arXiv preprint arXiv:1912.09877}\ } (\bibinfo {year} {2019})}\BibitemShut {NoStop}%
\bibitem [{\citenamefont {Erdmann}\ \emph {et~al.}(2018)\citenamefont {Erdmann}, \citenamefont {Glombitza},\ and\ \citenamefont {Walz}}]{erdmann2018deep}%
  \BibitemOpen
  \bibfield  {author} {\bibinfo {author} {\bibfnamefont {M.}~\bibnamefont {Erdmann}}, \bibinfo {author} {\bibfnamefont {J.}~\bibnamefont {Glombitza}},\ and\ \bibinfo {author} {\bibfnamefont {D.}~\bibnamefont {Walz}},\ }\bibfield  {title} {\bibinfo {title} {A deep learning-based reconstruction of cosmic ray-induced air showers},\ }\href@noop {} {\bibfield  {journal} {\bibinfo  {journal} {Astroparticle Physics}\ }\textbf {\bibinfo {volume} {97}},\ \bibinfo {pages} {46} (\bibinfo {year} {2018})}\BibitemShut {NoStop}%
\bibitem [{\citenamefont {Alvarez-Mu{\~n}iz}\ \emph {et~al.}(2024)\citenamefont {Alvarez-Mu{\~n}iz}, \citenamefont {Concei{\c{c}}{\~a}o}, \citenamefont {Costa}, \citenamefont {Gonz{\'a}lez}, \citenamefont {Pimenta},\ and\ \citenamefont {Tom{\'e}}}]{neutrinos_mercedes_2024}%
  \BibitemOpen
  \bibfield  {author} {\bibinfo {author} {\bibfnamefont {J.}~\bibnamefont {Alvarez-Mu{\~n}iz}}, \bibinfo {author} {\bibfnamefont {R.}~\bibnamefont {Concei{\c{c}}{\~a}o}}, \bibinfo {author} {\bibfnamefont {P.}~\bibnamefont {Costa}}, \bibinfo {author} {\bibfnamefont {B.}~\bibnamefont {Gonz{\'a}lez}}, \bibinfo {author} {\bibfnamefont {M.}~\bibnamefont {Pimenta}},\ and\ \bibinfo {author} {\bibfnamefont {B.}~\bibnamefont {Tom{\'e}}},\ }\bibfield  {title} {\bibinfo {title} {Potential of water-cherenkov air shower arrays for detecting transient sources of high-energy astrophysical neutrinos},\ }\href@noop {} {\bibfield  {journal} {\bibinfo  {journal} {Physical Review D}\ }\textbf {\bibinfo {volume} {110}},\ \bibinfo {pages} {023032} (\bibinfo {year} {2024})}\BibitemShut {NoStop}%
\bibitem [{\citenamefont {Abbasi}\ \emph {et~al.}(2023)\citenamefont {Abbasi} \emph {et~al.}}]{icecube2023galacticplane}%
  \BibitemOpen
  \bibfield  {author} {\bibinfo {author} {\bibfnamefont {R.}~\bibnamefont {Abbasi}} \emph {et~al.} (\bibinfo {collaboration} {IceCube}),\ }\bibfield  {title} {\bibinfo {title} {Observation of high-energy neutrinos from the galactic plane},\ }\href {https://doi.org/10.1126/science.adc9818} {\bibfield  {journal} {\bibinfo  {journal} {Science}\ }\textbf {\bibinfo {volume} {380}},\ \bibinfo {pages} {1338} (\bibinfo {year} {2023})},\ \Eprint {https://arxiv.org/abs/https://www.science.org/doi/pdf/10.1126/science.adc9818} {https://www.science.org/doi/pdf/10.1126/science.adc9818} \BibitemShut {NoStop}%
\bibitem [{\citenamefont {Aab}\ \emph {et~al.}(2021)\citenamefont {Aab} \emph {et~al.}}]{auger2021xmax}%
  \BibitemOpen
  \bibfield  {author} {\bibinfo {author} {\bibfnamefont {A.}~\bibnamefont {Aab}} \emph {et~al.} (\bibinfo {collaboration} {The Pierre Auger collaboration}),\ }\bibfield  {title} {\bibinfo {title} {Deep-learning based reconstruction of the shower maximum xmax using the water-cherenkov detectors of the pierre auger observatory},\ }\href {https://doi.org/10.1088/1748-0221/16/07/P07019} {\bibfield  {journal} {\bibinfo  {journal} {Journal of Instrumentation}\ }\textbf {\bibinfo {volume} {16}}\bibinfo  {number} { (07)},\ \bibinfo {pages} {P07019}}\BibitemShut {NoStop}%
\bibitem [{\citenamefont {Langer}\ \emph {et~al.}(2023)\citenamefont {Langer} \emph {et~al.}}]{Mass2023Auger}%
  \BibitemOpen
\bibfield  {number} {  }\bibfield  {author} {\bibinfo {author} {\bibfnamefont {N.}~\bibnamefont {Langer}} \emph {et~al.} (\bibinfo {collaboration} {The Pierre Auger collaboration}),\ }\bibfield  {title} {\bibinfo {title} {{Deep-Learning-Based Cosmic-Ray Mass Reconstruction Using the Water-Cherenkov and Scintillation Detectors of AugerPrime}},\ }\href {https://doi.org/10.22323/1.444.0371} {\bibfield  {journal} {\bibinfo  {journal} {PoS}\ }\textbf {\bibinfo {volume} {ICRC2023}},\ \bibinfo {pages} {371} (\bibinfo {year} {2023})}\BibitemShut {NoStop}%
\bibitem [{\citenamefont {Halim}\ \emph {et~al.}(2024)\citenamefont {Halim} \emph {et~al.}}]{auger2024mass}%
  \BibitemOpen
  \bibfield  {author} {\bibinfo {author} {\bibfnamefont {A.~A.}\ \bibnamefont {Halim}} \emph {et~al.} (\bibinfo {collaboration} {The Pierre Auger collaboration}),\ }\bibfield  {title} {\bibinfo {title} {Inference of the mass composition of cosmic rays with energies from $10^18.5$ to $10^20$ ev using the pierre auger observatory and deep learning},\ }\href@noop {} {\bibfield  {journal} {\bibinfo  {journal} {arXiv preprint arXiv:2406.06315}\ } (\bibinfo {year} {2024})}\BibitemShut {NoStop}%
\bibitem [{\citenamefont {Guest}\ \emph {et~al.}(2018)\citenamefont {Guest}, \citenamefont {Cranmer},\ and\ \citenamefont {Whiteson}}]{LHC_2018}%
  \BibitemOpen
  \bibfield  {author} {\bibinfo {author} {\bibfnamefont {D.}~\bibnamefont {Guest}}, \bibinfo {author} {\bibfnamefont {K.}~\bibnamefont {Cranmer}},\ and\ \bibinfo {author} {\bibfnamefont {D.}~\bibnamefont {Whiteson}},\ }\bibfield  {title} {\bibinfo {title} {Deep learning and its application to lhc physics},\ }\href {https://doi.org/10.1146/annurev-nucl-101917-021019} {\bibfield  {journal} {\bibinfo  {journal} {Annual Review of Nuclear and Particle Science}\ }\textbf {\bibinfo {volume} {68}},\ \bibinfo {pages} {161–181} (\bibinfo {year} {2018})}\BibitemShut {NoStop}%
\bibitem [{\citenamefont {Andrews}\ \emph {et~al.}(2020)\citenamefont {Andrews}, \citenamefont {Paulini}, \citenamefont {Gleyzer},\ and\ \citenamefont {Poczos}}]{LHC2020deep}%
  \BibitemOpen
  \bibfield  {author} {\bibinfo {author} {\bibfnamefont {M.}~\bibnamefont {Andrews}}, \bibinfo {author} {\bibfnamefont {M.}~\bibnamefont {Paulini}}, \bibinfo {author} {\bibfnamefont {S.}~\bibnamefont {Gleyzer}},\ and\ \bibinfo {author} {\bibfnamefont {B.}~\bibnamefont {Poczos}},\ }\bibfield  {title} {\bibinfo {title} {End-to-end physics event classification with cms open data: Applying image-based deep learning to detector data for the direct classification of collision events at the lhc},\ }\href@noop {} {\bibfield  {journal} {\bibinfo  {journal} {Computing and Software for Big Science}\ }\textbf {\bibinfo {volume} {4}},\ \bibinfo {pages} {1} (\bibinfo {year} {2020})}\BibitemShut {NoStop}%
\bibitem [{\citenamefont {Lv}\ \emph {et~al.}(2022)\citenamefont {Lv}, \citenamefont {Wang},\ and\ \citenamefont {Wu}}]{LHC2022deep}%
  \BibitemOpen
  \bibfield  {author} {\bibinfo {author} {\bibfnamefont {H.}~\bibnamefont {Lv}}, \bibinfo {author} {\bibfnamefont {D.}~\bibnamefont {Wang}},\ and\ \bibinfo {author} {\bibfnamefont {L.}~\bibnamefont {Wu}},\ }\bibfield  {title} {\bibinfo {title} {Deep learning jet images as a probe of light higgsino dark matter at the lhc},\ }\href@noop {} {\bibfield  {journal} {\bibinfo  {journal} {Physical Review D}\ }\textbf {\bibinfo {volume} {106}},\ \bibinfo {pages} {055008} (\bibinfo {year} {2022})}\BibitemShut {NoStop}%
\bibitem [{\citenamefont {Apolin{\'a}rio}\ \emph {et~al.}(2021)\citenamefont {Apolin{\'a}rio}, \citenamefont {Castro}, \citenamefont {Rom{\~a}o}, \citenamefont {Milhano}, \citenamefont {Pedro},\ and\ \citenamefont {Peres}}]{apolinario2021deep}%
  \BibitemOpen
  \bibfield  {author} {\bibinfo {author} {\bibfnamefont {L.}~\bibnamefont {Apolin{\'a}rio}}, \bibinfo {author} {\bibfnamefont {N.~F.}\ \bibnamefont {Castro}}, \bibinfo {author} {\bibfnamefont {M.~C.}\ \bibnamefont {Rom{\~a}o}}, \bibinfo {author} {\bibfnamefont {J.~G.}\ \bibnamefont {Milhano}}, \bibinfo {author} {\bibfnamefont {R.}~\bibnamefont {Pedro}},\ and\ \bibinfo {author} {\bibfnamefont {F.}~\bibnamefont {Peres}},\ }\bibfield  {title} {\bibinfo {title} {Deep learning for the classification of quenched jets},\ }\href@noop {} {\bibfield  {journal} {\bibinfo  {journal} {Journal of High Energy Physics}\ }\textbf {\bibinfo {volume} {2021}},\ \bibinfo {pages} {1} (\bibinfo {year} {2021})}\BibitemShut {NoStop}%
\bibitem [{\citenamefont {LeCun}\ \emph {et~al.}(2015)\citenamefont {LeCun}, \citenamefont {Bengio},\ and\ \citenamefont {Hinton}}]{deeplearning2015nature}%
  \BibitemOpen
  \bibfield  {author} {\bibinfo {author} {\bibfnamefont {Y.}~\bibnamefont {LeCun}}, \bibinfo {author} {\bibfnamefont {Y.}~\bibnamefont {Bengio}},\ and\ \bibinfo {author} {\bibfnamefont {G.}~\bibnamefont {Hinton}},\ }\bibfield  {title} {\bibinfo {title} {Deep learning},\ }\href@noop {} {\bibfield  {journal} {\bibinfo  {journal} {nature}\ }\textbf {\bibinfo {volume} {521}},\ \bibinfo {pages} {436} (\bibinfo {year} {2015})}\BibitemShut {NoStop}%
\bibitem [{\citenamefont {Vaswani}\ \emph {et~al.}(2017)\citenamefont {Vaswani}, \citenamefont {Shazeer}, \citenamefont {Parmar}, \citenamefont {Uszkoreit}, \citenamefont {Jones}, \citenamefont {Gomez}, \citenamefont {Kaiser},\ and\ \citenamefont {Polosukhin}}]{attention_2017}%
  \BibitemOpen
  \bibfield  {author} {\bibinfo {author} {\bibfnamefont {A.}~\bibnamefont {Vaswani}}, \bibinfo {author} {\bibfnamefont {N.}~\bibnamefont {Shazeer}}, \bibinfo {author} {\bibfnamefont {N.}~\bibnamefont {Parmar}}, \bibinfo {author} {\bibfnamefont {J.}~\bibnamefont {Uszkoreit}}, \bibinfo {author} {\bibfnamefont {L.}~\bibnamefont {Jones}}, \bibinfo {author} {\bibfnamefont {A.~N.}\ \bibnamefont {Gomez}}, \bibinfo {author} {\bibfnamefont {L.~u.}\ \bibnamefont {Kaiser}},\ and\ \bibinfo {author} {\bibfnamefont {I.}~\bibnamefont {Polosukhin}},\ }\bibfield  {title} {\bibinfo {title} {Attention is all you need},\ }in\ \href {https://proceedings.neurips.cc/paper_files/paper/2017/file/3f5ee243547dee91fbd053c1c4a845aa-Paper.pdf} {\emph {\bibinfo {booktitle} {Advances in Neural Information Processing Systems}}},\ Vol.~\bibinfo {volume} {30},\ \bibinfo {editor} {edited by\ \bibinfo {editor} {\bibfnamefont {I.}~\bibnamefont {Guyon}}, \bibinfo {editor} {\bibfnamefont {U.~V.}\ \bibnamefont {Luxburg}}, \bibinfo {editor}
  {\bibfnamefont {S.}~\bibnamefont {Bengio}}, \bibinfo {editor} {\bibfnamefont {H.}~\bibnamefont {Wallach}}, \bibinfo {editor} {\bibfnamefont {R.}~\bibnamefont {Fergus}}, \bibinfo {editor} {\bibfnamefont {S.}~\bibnamefont {Vishwanathan}},\ and\ \bibinfo {editor} {\bibfnamefont {R.}~\bibnamefont {Garnett}}}\ (\bibinfo  {publisher} {Curran Associates, Inc.},\ \bibinfo {year} {2017})\BibitemShut {NoStop}%
\bibitem [{\citenamefont {Watson}\ \emph {et~al.}(2023)\citenamefont {Watson} \emph {et~al.}}]{Watson2023HAWC}%
  \BibitemOpen
  \bibfield  {author} {\bibinfo {author} {\bibfnamefont {I.}~\bibnamefont {Watson}} \emph {et~al.} (\bibinfo {collaboration} {HAWC}),\ }\bibfield  {title} {\bibinfo {title} {{Deep Learning for the HAWC Observatory}},\ }\href {https://doi.org/10.22323/1.444.0927} {\bibfield  {journal} {\bibinfo  {journal} {PoS}\ }\textbf {\bibinfo {volume} {ICRC2023}},\ \bibinfo {pages} {927} (\bibinfo {year} {2023})}\BibitemShut {NoStop}%
\bibitem [{\citenamefont {Dosovitskiy}\ \emph {et~al.}(2020)\citenamefont {Dosovitskiy} \emph {et~al.}}]{ViT_2020}%
  \BibitemOpen
  \bibfield  {author} {\bibinfo {author} {\bibfnamefont {A.}~\bibnamefont {Dosovitskiy}} \emph {et~al.},\ }\bibfield  {title} {\bibinfo {title} {An image is worth 16x16 words: Transformers for image recognition at scale},\ }\href@noop {} {\bibfield  {journal} {\bibinfo  {journal} {arXiv preprint arXiv:2010.11929}\ } (\bibinfo {year} {2020})}\BibitemShut {NoStop}%
\bibitem [{\citenamefont {Heck}\ \emph {et~al.}(1998)\citenamefont {Heck}, \citenamefont {Knapp}, \citenamefont {Capdevielle}, \citenamefont {Schatz},\ and\ \citenamefont {Thouw}}]{CORSIKA}%
  \BibitemOpen
  \bibfield  {author} {\bibinfo {author} {\bibfnamefont {D.}~\bibnamefont {Heck}}, \bibinfo {author} {\bibfnamefont {J.}~\bibnamefont {Knapp}}, \bibinfo {author} {\bibfnamefont {J.}~\bibnamefont {Capdevielle}}, \bibinfo {author} {\bibfnamefont {G.}~\bibnamefont {Schatz}},\ and\ \bibinfo {author} {\bibfnamefont {T.}~\bibnamefont {Thouw}},\ }\bibfield  {title} {\bibinfo {title} {A monte carlo code to simulate extensive air showers},\ }\href@noop {} {\bibfield  {journal} {\bibinfo  {journal} {Report FZKA}\ }\textbf {\bibinfo {volume} {6019}} (\bibinfo {year} {1998})}\BibitemShut {NoStop}%
\bibitem [{\citenamefont {Ferrari}\ \emph {et~al.}(2005)\citenamefont {Ferrari}, \citenamefont {Ranft}, \citenamefont {Sala},\ and\ \citenamefont {Fass{\`o}}}]{fluka_2005}%
  \BibitemOpen
  \bibfield  {author} {\bibinfo {author} {\bibfnamefont {A.}~\bibnamefont {Ferrari}}, \bibinfo {author} {\bibfnamefont {J.}~\bibnamefont {Ranft}}, \bibinfo {author} {\bibfnamefont {P.~R.}\ \bibnamefont {Sala}},\ and\ \bibinfo {author} {\bibfnamefont {A.}~\bibnamefont {Fass{\`o}}},\ }\href@noop {} {\emph {\bibinfo {title} {FLUKA: A multi-particle transport code (Program version 2005)}}},\ \bibinfo {number} {CERN-2005-10}\ (\bibinfo  {publisher} {Cern},\ \bibinfo {year} {2005})\BibitemShut {NoStop}%
\bibitem [{\citenamefont {B{\"o}hlen}\ \emph {et~al.}(2014)\citenamefont {B{\"o}hlen}, \citenamefont {Cerutti}, \citenamefont {Chin}, \citenamefont {Fass{\`o}}, \citenamefont {Ferrari}, \citenamefont {Ortega}, \citenamefont {Mairani}, \citenamefont {Sala}, \citenamefont {Smirnov},\ and\ \citenamefont {Vlachoudis}}]{fluka_2014}%
  \BibitemOpen
  \bibfield  {author} {\bibinfo {author} {\bibfnamefont {T.}~\bibnamefont {B{\"o}hlen}}, \bibinfo {author} {\bibfnamefont {F.}~\bibnamefont {Cerutti}}, \bibinfo {author} {\bibfnamefont {M.}~\bibnamefont {Chin}}, \bibinfo {author} {\bibfnamefont {A.}~\bibnamefont {Fass{\`o}}}, \bibinfo {author} {\bibfnamefont {A.}~\bibnamefont {Ferrari}}, \bibinfo {author} {\bibfnamefont {P.~G.}\ \bibnamefont {Ortega}}, \bibinfo {author} {\bibfnamefont {A.}~\bibnamefont {Mairani}}, \bibinfo {author} {\bibfnamefont {P.~R.}\ \bibnamefont {Sala}}, \bibinfo {author} {\bibfnamefont {G.}~\bibnamefont {Smirnov}},\ and\ \bibinfo {author} {\bibfnamefont {V.}~\bibnamefont {Vlachoudis}},\ }\bibfield  {title} {\bibinfo {title} {The fluka code: developments and challenges for high energy and medical applications},\ }\href@noop {} {\bibfield  {journal} {\bibinfo  {journal} {Nuclear data sheets}\ }\textbf {\bibinfo {volume} {120}},\ \bibinfo {pages} {211} (\bibinfo {year} {2014})}\BibitemShut {NoStop}%
\bibitem [{\citenamefont {Ostapchenko}(2011)}]{QGSJET_2011}%
  \BibitemOpen
  \bibfield  {author} {\bibinfo {author} {\bibfnamefont {S.}~\bibnamefont {Ostapchenko}},\ }\bibfield  {title} {\bibinfo {title} {{Monte Carlo treatment of hadronic interactions in enhanced Pomeron scheme: I. QGSJET-II model}},\ }\href {https://doi.org/10.1103/PhysRevD.83.014018} {\bibfield  {journal} {\bibinfo  {journal} {Phys. Rev. D}\ }\textbf {\bibinfo {volume} {83}},\ \bibinfo {pages} {014018} (\bibinfo {year} {2011})},\ \Eprint {https://arxiv.org/abs/1010.1869} {arXiv:1010.1869 [hep-ph]} \BibitemShut {NoStop}%
\bibitem [{\citenamefont {Agostinelli}\ \emph {et~al.}(2003)\citenamefont {Agostinelli}, \citenamefont {Allison}, \citenamefont {Amako}, \citenamefont {Apostolakis}, \citenamefont {Araujo}, \citenamefont {Arce}, \citenamefont {Asai}, \citenamefont {Axen}, \citenamefont {Banerjee}, \citenamefont {Barrand} \emph {et~al.}}]{agostinelli2003geant4}%
  \BibitemOpen
  \bibfield  {author} {\bibinfo {author} {\bibfnamefont {S.}~\bibnamefont {Agostinelli}}, \bibinfo {author} {\bibfnamefont {J.}~\bibnamefont {Allison}}, \bibinfo {author} {\bibfnamefont {K.~a.}\ \bibnamefont {Amako}}, \bibinfo {author} {\bibfnamefont {J.}~\bibnamefont {Apostolakis}}, \bibinfo {author} {\bibfnamefont {H.}~\bibnamefont {Araujo}}, \bibinfo {author} {\bibfnamefont {P.}~\bibnamefont {Arce}}, \bibinfo {author} {\bibfnamefont {M.}~\bibnamefont {Asai}}, \bibinfo {author} {\bibfnamefont {D.}~\bibnamefont {Axen}}, \bibinfo {author} {\bibfnamefont {S.}~\bibnamefont {Banerjee}}, \bibinfo {author} {\bibfnamefont {G.~.}\ \bibnamefont {Barrand}}, \emph {et~al.},\ }\bibfield  {title} {\bibinfo {title} {Geant4—a simulation toolkit},\ }\href@noop {} {\bibfield  {journal} {\bibinfo  {journal} {Nuclear instruments and methods in physics research section A: Accelerators, Spectrometers, Detectors and Associated Equipment}\ }\textbf {\bibinfo {volume} {506}},\ \bibinfo {pages} {250} (\bibinfo {year}
  {2003})}\BibitemShut {NoStop}%
\bibitem [{\citenamefont {Allison}\ \emph {et~al.}(2006)\citenamefont {Allison} \emph {et~al.}}]{Geant4_2006}%
  \BibitemOpen
  \bibfield  {author} {\bibinfo {author} {\bibfnamefont {J.}~\bibnamefont {Allison}} \emph {et~al.},\ }\bibfield  {title} {\bibinfo {title} {Geant4 developments and applications},\ }\href@noop {} {\bibfield  {journal} {\bibinfo  {journal} {IEEE Transactions on Nuclear Science}\ }\textbf {\bibinfo {volume} {53 No. 1}},\ \bibinfo {pages} {270} (\bibinfo {year} {2006})}\BibitemShut {NoStop}%
\bibitem [{\citenamefont {Allison}\ \emph {et~al.}(2016)\citenamefont {Allison} \emph {et~al.}}]{Geant4_2016}%
  \BibitemOpen
  \bibfield  {author} {\bibinfo {author} {\bibfnamefont {J.}~\bibnamefont {Allison}} \emph {et~al.},\ }\bibfield  {title} {\bibinfo {title} {Recent developments in geant4},\ }\href@noop {} {\bibfield  {journal} {\bibinfo  {journal} {Nuclear Instruments and Methods in Physics Research A}\ }\textbf {\bibinfo {volume} {835}},\ \bibinfo {pages} {186} (\bibinfo {year} {2016})}\BibitemShut {NoStop}%
\bibitem [{Note1()}]{Note1}%
  \BibitemOpen
  \bibinfo {note} {This altitude corresponds to the altitude of the ALMA site in Chile.}\BibitemShut {Stop}%
\bibitem [{\citenamefont {O'Shea}\ and\ \citenamefont {Nash}(2015)}]{CNNs_2015_introduction}%
  \BibitemOpen
  \bibfield  {author} {\bibinfo {author} {\bibfnamefont {K.}~\bibnamefont {O'Shea}}\ and\ \bibinfo {author} {\bibfnamefont {R.}~\bibnamefont {Nash}},\ }\bibfield  {title} {\bibinfo {title} {An introduction to convolutional neural networks},\ }\href@noop {} {\bibfield  {journal} {\bibinfo  {journal} {arXiv preprint arXiv:1511.08458}\ } (\bibinfo {year} {2015})}\BibitemShut {NoStop}%
\bibitem [{\citenamefont {Dosovitskiy}\ \emph {et~al.}(2021)\citenamefont {Dosovitskiy} \emph {et~al.}}]{vit-base-patch16-224-in21k}%
  \BibitemOpen
  \bibfield  {author} {\bibinfo {author} {\bibfnamefont {A.}~\bibnamefont {Dosovitskiy}} \emph {et~al.},\ }\href@noop {} {\bibinfo {title} {{Hugging Face: vit-base-patch16-224-in21k.}}},\ \bibinfo {howpublished} {\url{https://huggingface.co/google/vit-base-patch16-224-in21k}} (\bibinfo {year} {2021}),\ \bibinfo {note} {[Online; accessed May-2023]}\BibitemShut {NoStop}%
\bibitem [{\citenamefont {Wu}\ \emph {et~al.}(2020)\citenamefont {Wu}, \citenamefont {Xu}, \citenamefont {Dai}, \citenamefont {Wan}, \citenamefont {Zhang}, \citenamefont {Yan}, \citenamefont {Tomizuka}, \citenamefont {Gonzalez}, \citenamefont {Keutzer},\ and\ \citenamefont {Vajda}}]{wu2020visual}%
  \BibitemOpen
  \bibfield  {author} {\bibinfo {author} {\bibfnamefont {B.}~\bibnamefont {Wu}}, \bibinfo {author} {\bibfnamefont {C.}~\bibnamefont {Xu}}, \bibinfo {author} {\bibfnamefont {X.}~\bibnamefont {Dai}}, \bibinfo {author} {\bibfnamefont {A.}~\bibnamefont {Wan}}, \bibinfo {author} {\bibfnamefont {P.}~\bibnamefont {Zhang}}, \bibinfo {author} {\bibfnamefont {Z.}~\bibnamefont {Yan}}, \bibinfo {author} {\bibfnamefont {M.}~\bibnamefont {Tomizuka}}, \bibinfo {author} {\bibfnamefont {J.}~\bibnamefont {Gonzalez}}, \bibinfo {author} {\bibfnamefont {K.}~\bibnamefont {Keutzer}},\ and\ \bibinfo {author} {\bibfnamefont {P.}~\bibnamefont {Vajda}},\ }\href@noop {} {\bibinfo {title} {Visual transformers: Token-based image representation and processing for computer vision}} (\bibinfo {year} {2020}),\ \Eprint {https://arxiv.org/abs/2006.03677} {arXiv:2006.03677 [cs.CV]} \BibitemShut {NoStop}%
\bibitem [{\citenamefont {Deng}\ \emph {et~al.}(2009)\citenamefont {Deng}, \citenamefont {Dong}, \citenamefont {Socher}, \citenamefont {Li}, \citenamefont {Li},\ and\ \citenamefont {Fei-Fei}}]{deng2009imagenet}%
  \BibitemOpen
  \bibfield  {author} {\bibinfo {author} {\bibfnamefont {J.}~\bibnamefont {Deng}}, \bibinfo {author} {\bibfnamefont {W.}~\bibnamefont {Dong}}, \bibinfo {author} {\bibfnamefont {R.}~\bibnamefont {Socher}}, \bibinfo {author} {\bibfnamefont {L.-J.}\ \bibnamefont {Li}}, \bibinfo {author} {\bibfnamefont {K.}~\bibnamefont {Li}},\ and\ \bibinfo {author} {\bibfnamefont {L.}~\bibnamefont {Fei-Fei}},\ }\bibfield  {title} {\bibinfo {title} {Imagenet: A large-scale hierarchical image database},\ }in\ \href@noop {} {\emph {\bibinfo {booktitle} {2009 IEEE conference on computer vision and pattern recognition}}}\ (\bibinfo {organization} {Ieee},\ \bibinfo {year} {2009})\ pp.\ \bibinfo {pages} {248--255}\BibitemShut {NoStop}%
\bibitem [{\citenamefont {Ridnik}\ \emph {et~al.}(2021)\citenamefont {Ridnik}, \citenamefont {Ben-Baruch}, \citenamefont {Noy},\ and\ \citenamefont {Zelnik-Manor}}]{Imagenet-21k_2021}%
  \BibitemOpen
  \bibfield  {author} {\bibinfo {author} {\bibfnamefont {T.}~\bibnamefont {Ridnik}}, \bibinfo {author} {\bibfnamefont {E.}~\bibnamefont {Ben-Baruch}}, \bibinfo {author} {\bibfnamefont {A.}~\bibnamefont {Noy}},\ and\ \bibinfo {author} {\bibfnamefont {L.}~\bibnamefont {Zelnik-Manor}},\ }\bibfield  {title} {\bibinfo {title} {Imagenet-21k pretraining for the masses},\ }\href@noop {} {\bibfield  {journal} {\bibinfo  {journal} {arXiv preprint arXiv:2104.10972}\ } (\bibinfo {year} {2021})}\BibitemShut {NoStop}%
\bibitem [{\citenamefont {Wolf}\ \emph {et~al.}(2019)\citenamefont {Wolf} \emph {et~al.}}]{huggingface_2019}%
  \BibitemOpen
  \bibfield  {author} {\bibinfo {author} {\bibfnamefont {T.}~\bibnamefont {Wolf}} \emph {et~al.},\ }\bibfield  {title} {\bibinfo {title} {Huggingface's transformers: State-of-the-art natural language processing},\ }\href@noop {} {\bibfield  {journal} {\bibinfo  {journal} {arXiv preprint arXiv:1910.03771}\ } (\bibinfo {year} {2019})}\BibitemShut {NoStop}%
\bibitem [{\citenamefont {Hunter}(2007)}]{matplotlib2007}%
  \BibitemOpen
  \bibfield  {author} {\bibinfo {author} {\bibfnamefont {J.~D.}\ \bibnamefont {Hunter}},\ }\bibfield  {title} {\bibinfo {title} {Matplotlib: A 2d graphics environment},\ }\href {https://doi.org/10.1109/MCSE.2007.55} {\bibfield  {journal} {\bibinfo  {journal} {Computing in Science \& Engineering}\ }\textbf {\bibinfo {volume} {9}},\ \bibinfo {pages} {90} (\bibinfo {year} {2007})}\BibitemShut {NoStop}%
\bibitem [{\citenamefont {Apolin{\'a}rio}\ \emph {et~al.}(2023)\citenamefont {Apolin{\'a}rio}, \citenamefont {Assis}, \citenamefont {Brogueira}, \citenamefont {Concei{\c{c}}{\~a}o}, \citenamefont {Costa}, \citenamefont {La~Mura}, \citenamefont {Pimenta},\ and\ \citenamefont {Tom{\'e}}}]{trigger_paper}%
  \BibitemOpen
  \bibfield  {author} {\bibinfo {author} {\bibfnamefont {L.}~\bibnamefont {Apolin{\'a}rio}}, \bibinfo {author} {\bibfnamefont {P.}~\bibnamefont {Assis}}, \bibinfo {author} {\bibfnamefont {P.}~\bibnamefont {Brogueira}}, \bibinfo {author} {\bibfnamefont {R.}~\bibnamefont {Concei{\c{c}}{\~a}o}}, \bibinfo {author} {\bibfnamefont {P.}~\bibnamefont {Costa}}, \bibinfo {author} {\bibfnamefont {G.}~\bibnamefont {La~Mura}}, \bibinfo {author} {\bibfnamefont {M.}~\bibnamefont {Pimenta}},\ and\ \bibinfo {author} {\bibfnamefont {B.}~\bibnamefont {Tom{\'e}}},\ }\bibfield  {title} {\bibinfo {title} {Identification of low energy neutral and charged cosmic ray events in large wide field observatorie},\ }\href@noop {} {\bibfield  {journal} {\bibinfo  {journal} {arXiv preprint arXiv:2310.15860}\ } (\bibinfo {year} {2023})}\BibitemShut {NoStop}%
\bibitem [{\citenamefont {Albert}\ \emph {et~al.}(2024)\citenamefont {Albert} \emph {et~al.}}]{hawc_ghsep_2024}%
  \BibitemOpen
  \bibfield  {author} {\bibinfo {author} {\bibfnamefont {A.}~\bibnamefont {Albert}} \emph {et~al.} (\bibinfo {collaboration} {HAWC}),\ }\bibfield  {title} {\bibinfo {title} {{Performance of the HAWC Observatory and TeV Gamma-Ray Measurements of the Crab Nebula with Improved Extensive Air Shower Reconstruction Algorithms}},\ }\href {https://doi.org/10.3847/1538-4357/ad5f2d} {\bibfield  {journal} {\bibinfo  {journal} {Astrophys. J.}\ }\textbf {\bibinfo {volume} {972}},\ \bibinfo {pages} {144} (\bibinfo {year} {2024})},\ \Eprint {https://arxiv.org/abs/2405.06050} {arXiv:2405.06050 [astro-ph.HE]} \BibitemShut {NoStop}%
\bibitem [{\citenamefont {Alfaro}\ \emph {et~al.}(2022)\citenamefont {Alfaro}, \citenamefont {Alvarez}, \citenamefont {{\'A}lvarez}, \citenamefont {Camacho}, \citenamefont {Arteaga-Vel{\'a}zquez}, \citenamefont {Rojas}, \citenamefont {Solares}, \citenamefont {Babu}, \citenamefont {Belmont-Moreno}, \citenamefont {Brisbois} \emph {et~al.}}]{hawc_ghsep_2022}%
  \BibitemOpen
  \bibfield  {author} {\bibinfo {author} {\bibfnamefont {R.}~\bibnamefont {Alfaro}}, \bibinfo {author} {\bibfnamefont {C.}~\bibnamefont {Alvarez}}, \bibinfo {author} {\bibfnamefont {J.}~\bibnamefont {{\'A}lvarez}}, \bibinfo {author} {\bibfnamefont {J.~A.}\ \bibnamefont {Camacho}}, \bibinfo {author} {\bibfnamefont {J.}~\bibnamefont {Arteaga-Vel{\'a}zquez}}, \bibinfo {author} {\bibfnamefont {D.~A.}\ \bibnamefont {Rojas}}, \bibinfo {author} {\bibfnamefont {H.~A.}\ \bibnamefont {Solares}}, \bibinfo {author} {\bibfnamefont {R.}~\bibnamefont {Babu}}, \bibinfo {author} {\bibfnamefont {E.}~\bibnamefont {Belmont-Moreno}}, \bibinfo {author} {\bibfnamefont {C.}~\bibnamefont {Brisbois}}, \emph {et~al.},\ }\bibfield  {title} {\bibinfo {title} {Gamma/hadron separation with the hawc observatory},\ }\href@noop {} {\bibfield  {journal} {\bibinfo  {journal} {Nuclear Instruments and Methods in Physics Research Section A: Accelerators, Spectrometers, Detectors and Associated Equipment}\ }\textbf {\bibinfo {volume} {1039}},\
  \bibinfo {pages} {166984} (\bibinfo {year} {2022})}\BibitemShut {NoStop}%
\bibitem [{\citenamefont {Atkins}\ \emph {et~al.}(2003)\citenamefont {Atkins}, \citenamefont {Benbow}, \citenamefont {Berley}, \citenamefont {Blaufuss}, \citenamefont {Bussons}, \citenamefont {Coyne}, \citenamefont {Delay}, \citenamefont {DeYoung}, \citenamefont {Dingus}, \citenamefont {Dorfan} \emph {et~al.}}]{milagro_compactness}%
  \BibitemOpen
  \bibfield  {author} {\bibinfo {author} {\bibfnamefont {R.}~\bibnamefont {Atkins}}, \bibinfo {author} {\bibfnamefont {W.}~\bibnamefont {Benbow}}, \bibinfo {author} {\bibfnamefont {D.}~\bibnamefont {Berley}}, \bibinfo {author} {\bibfnamefont {E.}~\bibnamefont {Blaufuss}}, \bibinfo {author} {\bibfnamefont {J.}~\bibnamefont {Bussons}}, \bibinfo {author} {\bibfnamefont {D.}~\bibnamefont {Coyne}}, \bibinfo {author} {\bibfnamefont {R.}~\bibnamefont {Delay}}, \bibinfo {author} {\bibfnamefont {T.}~\bibnamefont {DeYoung}}, \bibinfo {author} {\bibfnamefont {B.}~\bibnamefont {Dingus}}, \bibinfo {author} {\bibfnamefont {D.}~\bibnamefont {Dorfan}}, \emph {et~al.},\ }\bibfield  {title} {\bibinfo {title} {Observation of tev gamma rays from the crab nebula with milagro using a new background rejection technique},\ }\href@noop {} {\bibfield  {journal} {\bibinfo  {journal} {The Astrophysical Journal}\ }\textbf {\bibinfo {volume} {595}},\ \bibinfo {pages} {803} (\bibinfo {year} {2003})}\BibitemShut {NoStop}%
\end{thebibliography}%


\section*{Appendix}

\subsection{Optimal gamma selection efficiency}\label{sec:optimal_q}

\begin{figure}[htb]
 \centering
\includegraphics[width=0.8\linewidth]{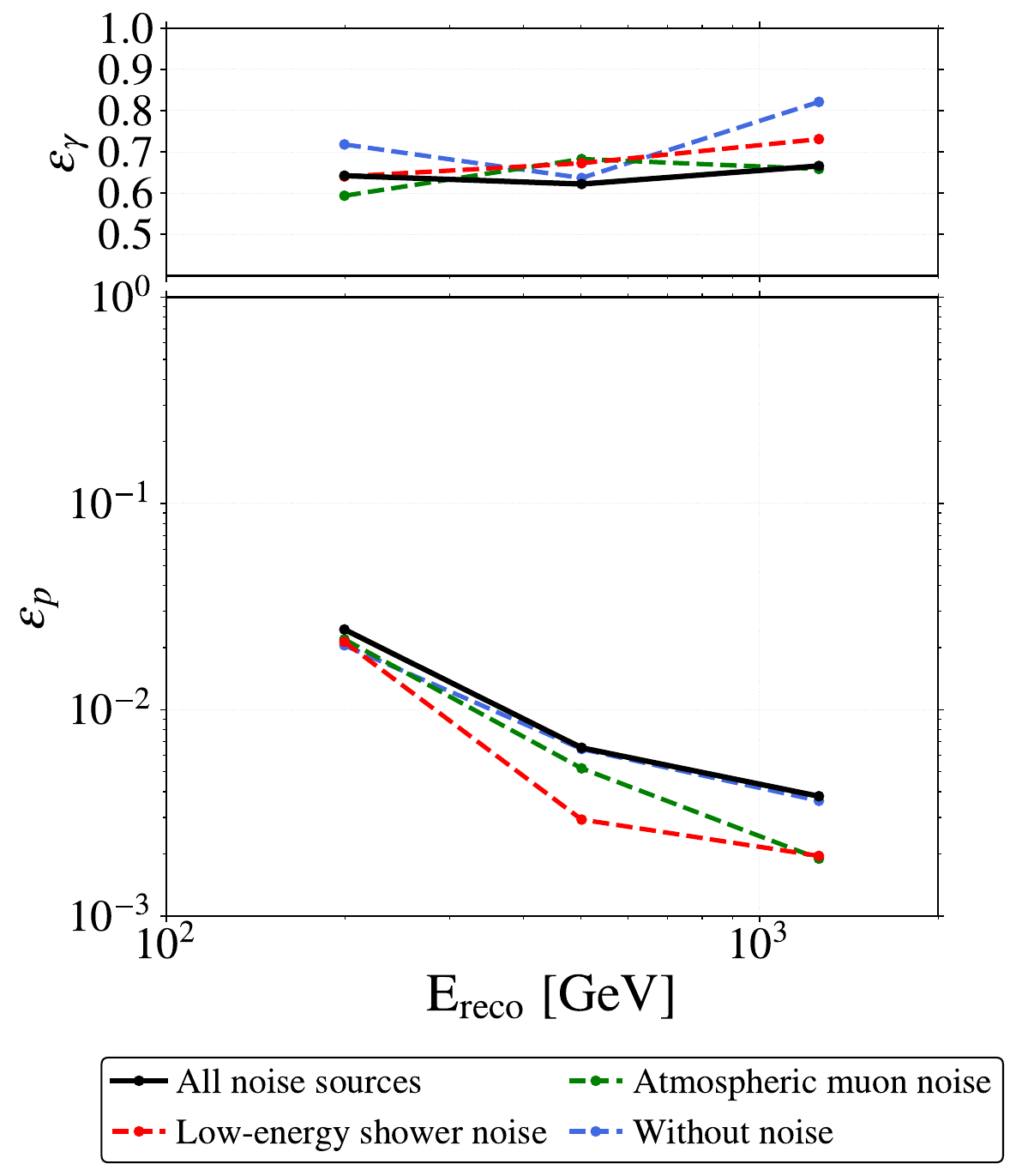}
\caption{The gamma (top) and proton (bottom) selection efficiency at different energies and noise sources, with the gamma selection efficiency set to achieve the optimal Q-factor.}
\label{fig:noise_results_best_q} 
\end{figure}

Figure \ref{fig:noise_results_best_q} presents the background rejection factor for a gamma efficiency selected to achieve the optimal Q-factor, as defined in Equation \ref{eq:q_factor}.
\begin{equation} \label{eq:q_factor}
    Q = \frac{\varepsilon_\gamma}{\sqrt{\varepsilon_p}}
\end{equation}

\subsection{Implementation of other gamma/hadron separators}\label{sec:hawc_gh_no_noise}

The principal HAWC gamma/hadron discriminators were adapted and incorporated into our simulation framework. Although HAWC typically calculates these variables using the charges at the PMT level, in this scenario, given the reduced size of the WCD, enhanced results were obtained by using the total charge from the three PMTs in the \textit{Mercedes} WCD. The MC core was used for the calculation of these discrimanators.

The first variable introduced was \textit{LIC}, which is the logarithmic transformation of the inverse of the \textit{compactness} parameter, an empirical parameter originally developed by the Milagro Collaboration \cite{hawc_ghsep_2022,milagro_compactness}. These variables were computed as follows:
\begin{equation}
\text { LIC }=\log _{10} \frac{1}{\text { compactness }}=\log _{10} \frac{\text { CxPE40 }}{N}
\end{equation}

where the \textit{CxPE40} represents the charge detected by the WCD located more than $40 \, \rm m$ from the shower core, which exhibits the highest effective charge, and \textit{N} is the total number of active stations. In stations with muons, it is anticipated that larger signals will be observed, consequently high LIC values are expected for showers initiated by hadrons.

Other discriminators can be obtained using the expected LDF of gamma rays, such as the \textit{PINCness} and the \textit{LDF} $\chi^2$. The \textit{PINCness} is a $\chi^2$-like measure of the smoothness of the charge footprint of the showers \cite{hawc_ghsep_2022}. To achieve this, as detailed in Equation (\ref{eq:PINCness}), it calculates the signal discrepancy between each WCD, $q_i$, and the average signal at stations equidistant from the shower core, $\left\langle\log _{10}\left(q_i\right)\right\rangle$, organising the stations into concentric rings with a width of $10 \, \rm m$. The discrepancy is normalised using the expected charge error $\sigma$, which was calculated for each energy bin using the stations within the gamma-ray events of the training datasets. Subsequently, a $\chi^2$-like metric is obtained by dividing the sum by the degrees of freedom, which equates to the count of stations, denoted as \textit{N}.
\begin{equation} \label{eq:PINCness}
\text { PINC }=\frac{1}{N} \sum_{i=0}^N \frac{\left(\log _{10}\left(q_i\right)-\left\langle\log _{10}\left(q_i\right)\right\rangle\right)^2}{\sigma^2}
\end{equation}
Alternatively, the \textit{LDF} $\chi^2$ employs a modified Nishimura-Kamata-Greisen (NKG) function \cite{hawc_ghsep_2024}, as described in Equation (\ref{eq:NKG}), to estimate the logarithm of the signal of a station, $\log _{10}\left(q_{\text {LDF }}\right)$, at a certain distance from the shower core, $r_{\rm core}$. Following this, the LDF $\chi^2$ is computed according to Equation (\ref{eq:LDFchi2}) utilizing HAWC's Moliere radius: $124.21 \, \rm m$.

\begin{equation} \label{eq:NKG}
    \begin{aligned}
\log _{10}(\text {NKG}) & =\log _{10} A+s\left(\log _{10}\left(\frac{r_{\rm core}}{124.21}\right)\right. \\
& \left.+\log _{10}\left(1+\frac{r_{\rm core}}{124.21}\right)\right) \\
& -3 \log _{10}\left(\frac{r_{\rm core}}{124.21}\right) \\
& -4.5 \log _{10}\left(1+\frac{r_{\rm core}}{124.21}\right)
\end{aligned}
\end{equation}

\begin{equation} \label{eq:LDFchi2}
\chi^2=\frac{1}{N}\sum_{i=0}^N \frac{\left(\log _{10}\left(q_i\right)-\log _{10}\left(q_{\text {LDF }}\right)\right)^2}{\sigma^2}
\end{equation}

\end{document}